\documentstyle[sprocl]{article}

\input{psfig}
\def\Journal#1#2#3#4{{#1} {\bf #2}, #3 (#4)}

\def\NPB{{\em Nucl. Phys.} B}
\def\PLB{{\em Phys. Lett.}  B}
\def\MPLA{{\em Mod. Phys. Lett.}  A}
\def\IJMP{{\em Int. J. Mod. Phys.}  A}
\def\PRL{\em Phys. Rev. Lett.}
\def\JHEP{\em J. High Energy Phys.}
\def\PRD{{\em Phys. Rev.} D}

\def\PR{\em Phys. Reports} 

\def\be{\begin{equation}}
\def\ee{\end{equation}}
\def\bea{\begin{eqnarray}}
\def\eea{\end{eqnarray}}

\begin{document}

\title{CONFINEMENT, MAGNETIC $Z_N$ SYMMETRY AND LOW ENERGY EFFECTIVE THEORY OF
GLUODYNAMICS}

\author{ A. KOVNER }

\address{Department of Physics, Theoretical Physics, 1 Keble Road,\\
Oxford OX1 3NP, England}

\maketitle\abstracts{In these notes I explain the idea how one could
understand confinement by studying the low energy effective dynamics
of non Abelian gauge theories. I argue that under some mild
assumptions,
the low energy dynamics is
determined universally by the spontaneous breaking of the magnetic symmetry
introduced by 't Hooft more than 20 years ago. The degrees of freedom
in the effective theory are magnetic vortices. They play a similar
role in confining dynamics to the role played by pions and sigma
in the chiral symmetry breaking dynamics. 
 I give explicit derivation
of the effective theory in 2+1 dimensional weekly coupled confining
models and give arguments that it remains qualitatively the same 
in strongly coupled 2+1 dimensional gluodynamics.
Confinement in this effective
theory is a very simple classical statement about long range
interaction between topological solitons, which follows by a simple
direct
classical calculation from
the structure of the effective Lagrangian.
I discuss the elements of this picture which generalize to 3+1
dimensions and point to the open questions still remaining.}

\section{Introduction.}
In these notes I am going to discuss 
an approach to understanding 
confinement in gluodynamics in terms of universal properties determined
by realization of global symmetries. 

First, let me be precise
in terminology. By gluodynamics I mean a non Abelian gauge theory without
dynamical fields in fundamental representation. Reason d'etre
being that I believe that confinement is the property of the pure glue
sector of a non Abelian theory and it is therefore important
to understand the pure glue theory. Presence of fundamental fermions 
in real life QCD is
from this point of view secondary and although it changes various
properties of the low energy dynamics in significant ways, it is
not driving the confinement phenomenon. This has been  
traditional 
point of view among the workers in field. It is however in marked 
contradiction to Gribovs ideas who maintained that it is the light 
fundamental fermions that bear ultimate responsibility for confinement.
These ideas, although very interesting will not be discussed in 
these notes.

Confinement is something of a mystery. It is certainly the most 
striking qualitative phenomenon in QCD. Still we do not even
have a satisfactory definition of what exactly is meant by this word.
Who is confined? Global color charges? No! States with nonvanishing
color charge (be it local or global) are not part of the Hilbert
space of the non Abelian gauge theory. The only physical states are 
those that satisfy Gauss' law - the color singlets. The global color 
is thus "confined" by construction, and that's certainly not what 
we mean. We are after a dynamical effect.
Consistently with the vagueness of the definition of the effect 
itself,
most models that have been put forward to explain it suffer from
similar blind spots.

Perhaps the most popular simple model of confinement 
is the dual superconductor model due to
't Hooft and Mandelstam ~\cite{dual}. 
The idea itself is very simple and as such certainly very appealing.
Consider a superconductor. It is described by the Landau - Ginzburg model
of the complex order parameter field $H$ (the Higgs field) coupled to the
vector potential $A_i$. The free energy (or Lagrangian of 
this Abelian Higgs model (AHM)
in the field
theoretical context) for static field configurations is
\begin{equation}
-L=|D_iH|^2+{1\over 2}B^2+\lambda (H^*H-v^2)^2
\end{equation}
where
\begin{equation}
D_i=\partial_i-igA_i
\end{equation}
The superconducting phase is characterized by a nonvanishing condensate of the
Higgs field
\begin{equation}
<H>=v
\end{equation}

The only way the magnetic field can penetrate a
superconductor is in the form of the flux tubes with the core size of the
penetration depth $l=(gv)^{-1}$ -
Abrikosov-Nielsen-Olesen (ANO) vortex. The magnetic flux inside a flux tube is
quantized in units of $2\pi/g$. The flux tube solution can be found numerically
and its properties in various limits are well understood analytically, see
for example ~\cite{vortex}.
Suppose now, that in addition to the charged field $H$ our model also has
very heavy magnetic monopoles. Since they are very heavy, they do
not affect the dynamics of the order parameter. But the structure of the vacuum
does affect strongly the interaction between the monopoles themselves. A
monopole and an antimonopole in the superconducting vacuum feel linear
confining potential. Since a monopole is a source of magnetic flux, this 
flux in the superconducting vacuum will form a ANO flux tube, which terminates on 
an antimonopole. The energy of the flux tube is proportional to its
length, hence the linear potential.

So, here is a theory in which certain objects are confined by a linear 
potential. 
The dual superconductor hypothesis asserts 
that the Yang Mills theory is basically an ``upside down'' version
of the Abelian Higgs model, as far as the confinement mechanism is concerned.
That is, if we use the following dictionary, the preceding discussion
describes confinement in any non Abelian gauge theory, which is in the 
confining phase.
The magnetic field of the Higgs model should be called the `''color electric 
field'', the magnetic monopoles should be called the ``color charges'', and by
analogy the electrically charged field $H$ should turn into
``color magnetic monopole''. 

This dual superconductor mechanism works in
the simplest known confining theory - 
Abelian compact $U(1)$ model.
This theory on the lattice is known to have a phase transition at a finite
value of the coupling constant. In the strongly coupled phase it is confining.
Note that in the Abelian, as opposed to non Abelian case, the global part
of the gauge group is a physical symmetry. Charged states do exist in the 
Hilbert space a priori, and in this case confinement
clearly means linear potential between these charges.
In this lattice theory one can
perform a duality transformation which transforms the compact $U(1)$
pure gauge theory into a noncompact $U(1)$ theory with charged fields, i.e.
Abelian Higgs model ~\cite{u1}. The ``Higgs'' field in the dual model
indeed 
describes
excitations carrying magnetic charge - magnetic monopoles. Those are condensed
in the confining phase and dual superconductivity captures perfectly the 
physics of confinement in this theory.

However as soon as one tries to generalize it to non Abelian theories many new
questions pop up.
In the Abelian theory the monopole charge is unambiguous and well defined. 
In non Abelian theories there is no gauge invariant
definition of a monopole, or for that matter even of the magnetic
charge 
that it carries.
In order to identify monopole - like objects, the common practice is to
to gauge fix the non Abelian gauge symmetry is a way that leaves only a $U(1)$
subgroup not fixed. This can be accomplished by gauge rotating any adjoint 
field $R$ into a canonical form. 
The original idea of 't Hooft was that the "monopoles"
should be identified with the singular obstructions to this gauge fixing, and the
physics should not depend on which particular quantity $R$ we use 
to set up the gauge fixing procedure.

This however turned out not to be the case. 
Since the definition of monopoles depends on the gauge, it is unavoidable 
that the correlation
between the quantitative properties of these monopoles (e.g. monopole density) 
and confinement
(the value of the string tension) also vary from gauge to gauge.
Although some qualitative properties are shared by monopoles 
in different gauges ~\cite{lucini}, the 
gauge dependence has been seen clearly in various lattice simulations. 
In the so called maximal Abelian gauge these correlations are
indeed very strong~\cite{maximal}. This gauge is designed to make the
gauge fixed Yang Mills Lagrangian to be as similar as possible to the compact 
$U(1)$ theory by 
attempting to maximally suppress the fluctuations of all fields except the 
color components 
of the gauge potential which belong to the Cartan subalgebra. In 
particular it looks like the monopoles contribute a large share of the string 
tension ~\cite{maximal}. The monopole
condensate disappears above the critical temperature in the deconfined phase. 
In other gauges correlations between the monopole properties and the confining 
properties of 
the theory are much weaker, and in some cases are completely absent~\cite{various}.
Even in the maximal Abelian gauge it is not clear that the monopole dominance 
is not a lattice
artifact. One expects that if the monopoles are indeed physical objects in the 
continuum limit,
their size would be of order of the only dimensional parameter in the theory 
$\Lambda_{QCD}^{-1}$.
The monopoles which appear in the context of lattice calculations on the other 
hand are all
pointlike objects.

During the recent years 
there have been some attempts to relate dual superconductivity
to confinement
somewhat more directly 
in continuum limit. The case in point is  
the exactly solvable
$N=2$ supersymmetric gauge theories ~\cite{seiwit}.
 Although $N=2$ super Yang Mills theory is
not confining, Witten and Seiberg showed that at one point in the moduli space
it contains massless monopoles. Introducing a particular supersymmetry breaking
perturbation the monopoles can be made to condense leading to confinement.
Unfortunately this construction does not address the genuine non Abelian
dynamical questions. The $N=2$ super Yang Mills theory close to the monopole
point is essentially Abelian. The spectrum contains an Abelian gauge field,
while the analogs of the non Abelian gluons are massive due to the
Higgs mechanism. Confinement in this model is therefore similar to the Abelian 
confinement of the compact Abelian theory. As discussed in ~\cite{yung}
the properties of the spectrum and various characteristics of
confining 
strings are
very different in this model and in nonsupersymmetric QCD. 
In particular it is not known whether the Abelian monopole charge
survives 
at all
in the genuine non Abelian regime and whether the condensation of
Abelian 
monopoles
is at all relevant to confinement in the non Abelian limit.
It is entirely unclear 
at this point how this supersymmetry inspired construction can help us
understand
the real theory of strong interactions.

Thus although the dual superconductor picture is appealing due to its
simplicity, it is far from having been substantiated.
To my mind there is quite a general objection to it.
The dual superconductor 
hypothesis in effect states
that the effective low energy theory of the pure Yang Mills theory
when written in terms of appropriate field variables is an Abelian
Higgs 
model. 
If so, the low lying spectrum of pure Yang Mills 
should be the same as that of an 
Abelian 
Higgs model.

A statement about the form of the effective Lagrangian is not easily 
verified, since it 
necessitates an appropriate choice of variables. On the other hand a statement 
about the spectrum
does not depend on the choice of variables and is in this sense universal and 
easily verifiable.
The spectrum of the Abelian Higgs model is well known. The two lowest
mass 
excitations 
are a
massive vector particle and a massive scalar particle. The 
role of the vector particle is 
crucial in the framework of the dual superconductor model.
Indeed in the Seiberg-Witten model the light vector is present in the spectrum.
However this is not the case in gluodynamics.
The spectrum of the Yang Mills theory is not known from analytical calculations. However, in 
recent years a rather clear picture of it emerged from the lattice
simulations ~\cite{glueballs}.
The lowest lying particle in the spectrum is a scalar glueball with the mass $1.5-1.7$ 
Gev.
The second excitation is a spin 2 tensor glueball with a mass of around $2.2$ Gev.
Vector (and pseudo vector) glueballs are conspicuously missing in the lowest lying part
of the spectrum. The simulations indicate that they are relatively
heavy 
\footnote{Qualitatively,  precisely
this  pattern was predicted
20 years ago~\cite{N} on the basis of QCD low-energy theorems.}
with masses above
2.8 Gev ~\cite{glueballs}. The pattern of 
the Yang Mills spectrum therefore seems to be rather different from the
one suggested by the Abelian Higgs model. Since the vector glueball is so 
much heavier
than the scalar and tensor ones, it seems quite unlikely that it plays so prominent 
a role
in the confinement mechanism as the one reserved for the massive photon in the dual
superconductor scenario.

Of course, it's not over until it's over. Since no dynamical solution
to the problem exist we cannot completely rule out the possibility
that it is the properties of the high energy, rather than of 
low energy part of the spectrum,
that are directly related to confinement. To me however this sounds
like a far cry.
It seems much more likely that the key is at low
energies.
And so rather than looking for it under the lamp post of the dual
superconductor model,
we may, while we 
are at it, ask a more general question. 
Can we learn about the mechanism of confinement
by studying the effective low energy theory? 
This does not sound like such a stupid idea.
If confinement (whatever it is) is a low energy phenomenon, it is
very likely that inside the low energy effective theory 
we would be able to understand it in a simple way.
The catch is of course obvious: we don't know what the low energy
theory is and what principles determine it.
And so it could be that determining the low energy theory is itself 
even more difficult an undertaking than just understanding confinement.

In these notes I will try to argue that this is not necessarily the
case, 
and that such an undertaking is not as hopeless as it may seem.
There are instances when we can understand main features
of the low energy dynamics without being able to solve in detail
the ``microscopic'' theory. This happens whenever we are lucky enough
to have a spontaneously broken global continuous symmetry. The Goldstone
theorem assures us that such a spontaneous symmetry breaking 
results in appearance of massless particles. Those of course are natural
low energy degrees of freedom. Moreover the 
original symmetry must be manifest in the effective
theory that governs the interactions between the Goldstone bosons,
albeit
its implementation is nonlinear. This severely constrains 
the interactions of the Goldstone bosons and in fact 
lends considerable predictive power to the effective theory.
The classic example of this type is the spontaneous breaking of chiral
symmetry which has the effect that the low energy physics is dominated
by pions and is described by the chiral effective Lagrangian
~\cite{chiral}. 
In real life the chiral symmetry is broken
explicitly and the pion is massive. But effects of small 
explicit breaking are easily accommodated in the chiral Lagrangian.
The great thing about the chiral Lagrangian of course is that it is
universal.
It does not care what exact dynamics is responsible for the symmetry
breaking, what where the degrees of freedom of ``microscopic'' theory
or any other fine details. All you need to know is that there was a
symmetry, and this symmetry was spontaneously broken.

Can some universal considerations of a similar kind
determine the structure of low energy theory in pure
gluodynamics?
The chiral symmetry here is certainly of no relevance. Nevertheless I
will argue that the symmetry path is a very fruitful one. The main
thesis is this. There is a global discreet symmetry in gluodynamics,
which is spontaneously broken in the vacuum. 
In the following this symmetry will be called the magnetic $Z_N$.
Although the symmetry is
discreet and therefore does not have all the bliss of the
Goldstone theorem, under some natural assumptions it does indeed
determine the low energy dynamics. In physical 3+1 dimensional theory
this symmetry is of a somewhat unusual type - its charge is not a volume
integral, but rather a surface integral. The implications of such a
symmetry for the low energy Lagrangian have to be studied in more
detail and this has not been done so far. However one can go much further
in 2+1 dimensions, where this symmetry is of a familiar garden
variety. The larger part of these notes will be devoted therefore to
2+1 dimensional non Abelian theories. 
The structure of these notes is the following. Sections 2-5 are
devoted to 2+1 dimensional physics. In section 2
I will discuss in detail the
nature of the magnetic $Z_N$ symmetry and the explicit construction of
both, the generator of the group and the local order parameter.
In Section 3 I will show how this symmetry is realized in different
phases of Abelian gauge theories. I will show that in the Coulomb
phase this symmetry is spontaneously broken and in the Higgs phase it
is
respected by the vacuum state. In the Abelian theories the magnetic
symmetry is continuous $U(1)$ group, and so its spontaneous breaking
leads to the appearance of a massless particle - the photon.
I will derive the low energy effective
Lagrangian that describes this symmetry breaking pattern and show that
it exhibits logarithmic confinement in the Coulomb phase.
In Section 4 I will discuss weakly interacting confining
theories, like the Georgi-Glashow model. Here the magnetic symmetry is
discreet, but it is also spontaneously broken like in the Coulomb
phase of QED. Again it is possible to derive 
the effective Lagrangian which
follows from this symmetry breaking pattern. I will demonstrate
that due to the fact that the magnetic symmetry is discreet
the effective theory exhibits linear confinement, and that this 
confinement mechanism in the effective theory appears very simply on
the classical level.
In Section 5  I will give arguments that the basic structure
of the effective Lagrangian as well as the physics of confinement
stays the same in the pure gluodynamics limit. I will discuss the
similarities and differences between the confining properties of the
weakly interacting non Abelian theories and gluodynamics.
In section 6 I will discuss the generalization of these ideas to the
physical dimensionality of 3+1, including the symmetry structure, the
order parameter and the realization of the symmetry. The explicit
construction of the low energy effective theory in 3+1 dimensions has
not been completed, although I will discuss some modest steps in 
this direction.
Finally section 7 is devoted to the discussion of the lessons, past
and future that can be hopefully learned from this approach.

\section{The magnetic symmetry in 2+1 dimensions.}

A while back 't Hooft gave an argument establishing
that a non Abelian $SU(N)$ gauge theory 
without charged fields in fundamental representation
possesses a global $Z_N$ symmetry ~\cite{thooft}. 

Consider first a theory with several adjoint Higgs fields so that varying
parameters in the Higgs sector 
the $SU(N)$ gauge symmetry can be broken completely. In
this phase the perturbative spectrum will contain the usual massive
``gluons'' and Higgs particles. However in addition to those there will
be heavy stable magnetic vortices. Those are the analogs of
Abrikosov-Nielsen-Olesen vortices in the superconductours and they
must be stable by virtue of the following topological argument.
The vortex configuration away from the vortex core has all the fields
in the pure gauge configuration
\begin{equation}
H^{\alpha}(x)=U(x)h^{\alpha}, \ \ \ \ A^\mu=iU\partial^\mu
U^\dagger
\end{equation}
Here the index $\alpha$ labels the scalar fields in the theory,
$h^\alpha$ are the constant vacuum expectation values of these fields, and
$U(x)$ is a unitary matrix. As one goes around the location of
the vortex in space, the matrix $U$ winds nontrivially in the gauge
group. This is possible, since the gauge group in the theory without
fundamental fields is $SU(N)/Z_N$ and it has a nonvanishing first
homotopy group $\Pi_1(SU(N)/Z_N)=Z_N$. Practically it means that when
going full circle around the vortex location, $U$  does not
return to
the same $SU(N)$ group element $U_0$, but rather ends up at
$\exp\{i{2\pi\over N}\}U_0$. Adjoint fields do not feel this type of
discontinuity in $U$ and therefore the energy of such a configuration
is finite. Since such a configuration can not be smoothly deformed
into a trivial one, a single vortex is stable. Processes involving
annihilation
of N such
vortices into vacuum are allowed since N-vortex configurations 
are topologically
trivial.
One can of course find explicit vortex solutions once the Higgs
potential is specified. 
As any other semi classical solution in the weak coupling
limit the energy of such a vortex is inversely proportional to the
gauge coupling constant and therefore very large.
One is therefore faced with the situation where the spectrum of the
theory
contains a stable particle even though its mass is much higher than
masses of many other particles (gauge and Higgs bosons) 
and the phase space for its decay into
these particles is enormous. The only possible reason for the existence of
such a heavy stable particle is that it  carries a conserved quantum
number.
The theory therefore must possess a global symmetry which is unbroken
in the completely Higgsed phase. The symmetry group must be $Z_N$
since the number of vortices is only conserved modulo $N$.
't Hooft dubbed this symmetry ``topological'', but I prefer to call
it ``magnetic'' for reasons that will become apparent in a short while.

Now imagine changing smoothly the parameters in the Higgs sector so
that the expectation values of the Higgs fields become smaller and
smaller, and finally the theory undergoes a phase transition into the
confining phase. One can further change the parameters so that the
adjoint scalars become heavy and eventually decouple completely from
the glue. This limiting process does not change the topology of the gauge
group and therefore does not change the symmetry content of the
theory. One concludes that the pure Yang-Mills theory also
possesses a $Z_N$ symmetry. 
Of course since the confining phase is separated from the completely
Higgsed phase by a phase transition one may expect that the $Z_N$
symmetry in the confining phase is realized differently in the
confining vacuum 
than in the
completely "Higgsed" phase. In
fact
the original paper of 't Hooft as well as subsequent work~\cite{zn}
convincingly 
argued that in the confining phase the $Z_N$ symmetry is spontaneously
broken and this breaking is related to the confinement phenomenon.
We will have more to say about this later.

The physical considerations given above can be put on a firmer formal
basis. In particular one can construct explicitly the generator of the $Z_N$
as well as the order parameter associated with it 
- the operator that creates the
magnetic vortex ~\cite{kovner1}.
We start this discussion by considering Abelian theories, where things
are simpler and are completely under control.

\subsection{Abelian theories.}

In the $U(1)$  case the homotopy group is $Z$ and so 
the magnetic symmetry is $U(1)$ rather than $Z_N$. 
It is in fact absolutely
straightforward to identify the relevant charge. It is none other than
the magnetic flux through the equal time plane, with the associated
conserved current being the dual of the electromagnetic field strength
\begin{equation}
\Phi=\int d^2x B(x),\ \ \ \ \ \ \ \ \ \ \partial^\mu\tilde F_\mu=0
\label{phiqed}
\end{equation}

The current conservation is insured by the Bianchi identity.
It may come as a surprise that we are considering seriously a current
whose
conservation equation is an ``identity''. However ``identity'' is a
relative thing. The conservation equation is trivial only because we have
written the
components of the field strength tensor in terms of the vector
potential. But the introduction of vector
potential is nothing but potentiation of the conserved current $\tilde
F_\mu$,
that is explicit solution of the conservation equation. In exactly 
the same way
one can potentiate any conserved current, and such a potentiation will
turn the pertinent conservation equation into identity. Thus $\tilde
F_\mu$
has exactly the same status as any other 
local conserved gauge invariant current, and should be treated as
such.

Once we have the current and the charge, we also know the elements of the
symmetry group.
A group element of the $U(1)$ magnetic symmetry group is
\begin{equation}
W_\alpha(\infty)=\exp\{i\alpha\Phi\}
\end{equation}
for arbitrary value of $\alpha$.
The notation $W$ is not accidental here, since the group element is 
indeed a large spatial Wilson loop
defined on a contour that encloses the whole system.

The question that might bother us is whether this group
acts at all nontrivially on any local physical observable in our theory.
The obvious gauge invariant observables like $B$ and $E$ commute
with $W$.
There is however another set of local gauge invariant observables in
the theory which do indeed transform nontrivially under the action of $W$.
Consider, following 't Hooft the operator  of the "singular gauge
transformation"
\begin{equation}
V(x)=\exp {\frac{i}{g}}\int d^2y\
\left[\epsilon_{ij}{\frac{(x-y)_j}{(x-y)^2}}
E_i(y)+\Theta(x-y)J_0(y)\right]  
\label{vqed1}
\end{equation}
where $\Theta(x-y)$ is the polar angle function and $J_0$ is the
electric 
charge density of whatever matter fields
are present in the theory. 
The cut discontinuity in the function $\Theta$
looks bothersome, but is in fact not physical and
completely harmless. The gauge function jumps across the discontinuity
by $2\pi/g$, but since the only dynamical fields in the theory have
charges
that are integer multiples of $g$ 
the discontinuity is not observable. 
The cut
 can be chosen parallel
to the horizontal axis.
Using the Gauss' law constraint this can be cast in a different form,
which we will find more convenient for our discussion
\begin{equation}
V(x)=\exp {\frac{2\pi i}{g}}\int_C dy^i\
\epsilon_{ij}E_i(y)
\label{vqed}
\end{equation}
where the integration goes along the cut of the function $\Theta$
which starts at the point $x$ and goes to spatial infinity. 
In this form it is clear
that the
operator does not depend on where precisely one chooses the 
cut to lie. 
To see this, note that 
changing the position of the cut $C$ to $C'$ adds to the phase 
${2\pi\over g}\int_{S}d^2x \partial_iE^i$ where $S$ is the area
bounded 
by $C-C'$. 
In the
theory we consider 
only charged particles with charges multiples of $g$ are present. Therefore 
the charge within any closed area is a
multiple integer of the gauge coupling $ \int_{S}d^2x \partial_iE^i=gn$
and the extra phase factor is always unity.
The only point in space where the action of $V(x)$ on any physical state
is nontrivial is the point $x$.
The field $V(x)$ therefore acts like any other local field. With a little
more work one can prove not only that $V(x)$ is a local field, in the
sense that it commutes with any other local gauge invariant operator
$O(y)$,  $x\ne y$ but also that it is a bona fide Lorentz scalar.
~\cite{kovner2}.

The physical meaning of the operator $V$ is very simple. Calculating
its commutator with the magnetic field $B$ we find
\begin{equation}
V(x)B(y)V^\dagger(x)=B(y)+{2\pi\over g}\delta^2(x-y)
\end{equation}
Thus
$V$ creates a pointlike magnetic vortex of flux
$2\pi/g$.
This commutator also tells us that
\begin{equation}
W^\dagger_{\alpha}V(x)W_\alpha=e^{i{2\pi\alpha\over g}}V(x)
\end{equation}
and so $V$ is indeed the local eigenoperator of the magnetic $U(1)$
symmetry group.

Eqs.(\ref{phiqed},\ref{vqed}) formalize the physical arguments of 't Hooft in
the Abelian case.
We have explicitly constructed the generator and the local order
parameter of the magnetic symmetry.

\subsection{Non-Abelian theories at weak coupling.}

Let us now move onto the analogous construction for non Abelian theories.
Ultimately we are
interested in the pure Yang - Mills theory. It is however                 
illuminating to start with the theory with an adjoint Higgs field and
take the decoupling limit explicitly later. For simplicity we
discuss the $SU(2)$ gauge theory.
Consider the
Georgi-Glashow model - $SU(2)$ gauge theory with an adjoint Higgs
field.
\begin{equation}
{\cal L}=-\frac{1}{4}F_{\mu \nu }^{a}F^{a\mu \nu }+\frac{1}{2}({\cal D}_{\mu
}^{ab}H^{b})^{2}+\tilde{\mu}^{2}H^{2}-\tilde{\lambda}(H^{2})^{2}  \label{lgg}
\end{equation}
where 
\begin{equation}
{\cal D}_{\mu }^{ab}H^{b}=\partial _{\mu }H^{a}-gf^{abc}A_{\mu }^{b}H^{c}
\end{equation}

At large and positive $\tilde{\mu}^{2}$ the model is weakly coupled. 
The $SU(2)$ gauge symmetry is broken down
to $U(1)$ and the Higgs mechanism takes place. Two gauge bosons, $W^{\pm }$,
acquire a mass, while the third one, the ``photon'', remains massless
to all orders in perturbation theory. 
The theory in this region of parameter space resembles very
much electrodynamics with vector charged fields.
The Abelian construction can therefore be repeated.
The $SU(2)$ gauge invariant analog of the conserved dual field strength is
\begin{equation}
\tilde{F}^{\mu }={1\over 2}[\epsilon_{\mu\nu\lambda}F^a_{\nu\lambda}n^a-
\frac{1}{g}\epsilon ^{\mu \nu \lambda}
\epsilon ^{abc}n_{a}({\cal D}_{\nu }n)^{b}({\cal D}_{\lambda }n)^{c} ]
 \label{F}
\end{equation}
where $n^a\equiv{\frac{H^a}{|H|}}$ is the unit vector in the
direction of the Higgs field.
Classically
this current satisfies the conservation equation
\begin{equation}
\partial^\mu\tilde F_\mu=0
\label{f}
\end{equation}
The easiest way to see this is to choose a unitary gauge of the form 
$H^a(x)=H(x)\delta^{a3}$. In this gauge $\tilde F$ is equal to the
Abelian part of the dual field strength in the third direction in
color space. Its conservation then follows by the Bianchi
identity. 
Thus classically the theory has a conserved $U(1)$ magnetic charge
$\Phi=\int d^2x \tilde F_0$ just like QED.
However the unitary gauge can not be imposed at the points where $H$
vanishes,
which necessarily happens in the core of an 't Hooft-Polyakov
monopole. It is well known of course ~\cite{polyakov} that the
monopoles are the
most important nonperturbative configurations in this model. Their presence
leads to a nonvanishing small mass for the photon
as well as
to confinement of the charged gauge bosons with a tiny
nonperturbative string tension.
As far as the monopole effects on the magnetic flux, 
their presence leads to a
quantum anomaly in the conservation equation (\ref{f}). As a result only
the discrete $Z_2$ subgroup
of the transformation group generated by $\Phi$
remains
unbroken in the quantum theory. The detailed discussion of this
anomaly, the residual $Z_2$ symmetry  and their 
relation to monopoles is given in ~\cite{kovner1}. 
The nonanomalous $Z_2$ magnetic
symmetry transformation is generated by the operator 
\begin{equation}
U=\exp\{i{g\over 2}\Phi\}
\label{u}
\end{equation}
The order parameter for the magnetic $Z_2$ symmetry is constructed
analogously to QED as a singular gauge transformation generated by the
 gauge invariant electric charge operator
\begin{equation}
J^\mu=\epsilon^{\mu\nu\lambda}\partial_\nu (\tilde F^a_{\lambda}n^a), 
\ \ \ \ Q=\int d^2x J_0(x)  
\label{qqcd}
\end{equation}
Explicitly
\begin{eqnarray}
\nonumber
V(x)&=&\exp {\frac{i}{g}}\int d^2y\ \left[\epsilon_{ij}{\frac{(x-y)_j}{(x-y)^2}
} n^a(y)E^a_i(y)+\Theta(x-y)J_0(y)\right]  \\
&=&\exp {\frac{2\pi i}{g}}\int_C dy^i
\epsilon_{ij}n^aE^a_j(y)
\label{VQCD}
\end{eqnarray}
One can think of it as 
a singular $SU(2)$ gauge transformation with the field
dependent gauge function 
\begin{equation}
\lambda^a(y)={\frac{1}{g}}\Theta(x-y)n^a(\vec y)  
\label{lambda}
\end{equation}
This field dependence of the gauge function ensures the gauge 
invariance of the operator $V$.
Just like in QED it can be shown~\cite{kovner1}, ~\cite{kovner2} that
the operator $V$ is a local scalar field.
Again like in QED, the vortex operator $V$ is a local eigenoperator of
the Abelian magnetic field $B(x)=\tilde{F}_{0}$. 
\begin{equation}
\lbrack V(x),B(y)\rbrack=-{\frac{2\pi }{g}}V(x)\delta ^{2}(x-y)  \label{com}
\end{equation}
That is to say, when acting on a state it creates a pointlike magnetic vortex
which carries a quantized unit of magnetic flux. 

The magnetic $Z_2$ acts
on the vortex field $V$ as a phase rotation by $\pi$
\begin{equation}
e^{i{g\over 2} \Phi }V(x)e^{-i{g\over 2} \Phi }=-V(x)
\label{magntr}
\end{equation}

This is the explicit realization of the magnetic $Z_2$ symmetry in 
the Georgi-Glashow model. 

\subsection{Gluodynamics.}

From the Georgi Glashow model we can get easily to the pure Yang
Mills theory.
This is achieved by smoothly varying the 
$\tilde\mu^2$ coefficient in the Lagrangian so that it 
becomes negative and eventually arbitrarily large.
In this limit the Higgs field has a large mass and therefore decouples
leaving the pure gluodynamics behind.
It is well known that
in this model
the weakly coupled Higgs regime and strongly coupled
confining regime are not separated by a phase 
transition~\cite{fradkin}. 
The pure Yang Mills limit in this model is therefore smooth.

In the pure Yang Mills limit the expressions Eq.~\ref{F},\ref{VQCD},\ref{u}
have to be taken with care. 
When the mass of the Higgs field is very large, the 
configurations 
that dominate the path integral of the theory 
are those with very small value of the
modulus of
the Higgs field $|H|\propto 1/M$.
The modulus of the Higgs field in turn controls the fluctuations of
the unit vector $n^a$,
since the kinetic term for $n$ in the Lagrangian is $|H|^2(D_\mu n)^2$.
Thus as the mass of the Higgs field increases the fluctuations of $n$
grow in both, amplitude and frequency and the magnetic field operator $B$ as defined in
Eq.~\ref{F} fluctuates wildly.
This situation is of course not unusual. It happens whenever one
wants to consider in the effective low energy theory an operator which
explicitly depends on fast, high energy variables. The standard way to deal
with it is to integrate over the fast variables. 
There could be two possible outcomes of this integrating out
procedure. Either the
operator in question becomes trivial (if it depends strongly on the
fast variables), or its reduced version is well defined and regular on
the low energy Hilbert space.
The ``magnetic field'' operator $B$ in Eq.~\ref{F} is obviously of
the first type. Since in the pure Yang Mills limit all 
the orientations of $n^a$ are equally probable, integrating over the
Higgs field at fixed $A_\mu$ will lead to vanishing of $B$. However
what interests us is not so much the magnetic field but rather the
generator of the magnetic $Z_2$ transformation. It is actually
instructive to consider the operator that performs the $Z_2$
transformation not everywhere in space, but only inside a contour $C$
\be
U_C=\exp \{i{g\over 2}\int_S d^2xB(x)\}
\label{uc}
\ee
with the area $S$ being bounded by $C$.
In the limit of gluodynamics we are lead to
consider the operator
\begin{eqnarray}
U_C=lim_{H\rightarrow 0}&&\int Dn^a\exp\Big\{-|H|^2(\vec D n_a)^2\Big\}\\
&&\exp\Big\{i{g\over 4}\int_C d^2x\big(\epsilon_{ij}F^a_{ij}n^a-
\frac{1}{g}\epsilon ^{ij}\epsilon ^{abc}n_{a}({\cal D}_{i }n)^{b}
({\cal D}_{j }n)^{c}\big ) \Big\}\nonumber
\label{uc1}
\end{eqnarray}
The weight for the integration over $n$ is the 
kinetic term for the isovector $n_a$. As was noted
before the action does not depend on $n^a$ in the YM limit since
$H^2\rightarrow 0$.
The first term in Eq.~\ref{uc1} 
however regulates the path integral and we keep it for this reason.
This operator
may look somewhat unfamiliar at first sight. However in a remarkable
paper ~\cite{dp} Diakonov and Petrov showed that Eq.~\ref{uc1} is equal to 
the trace of the fundamental Wilson loop along the contour 
$C$\footnote{We note that Diakonov and Petrov had to introduce a regulator
to define the path integral over $n$. The regulator they required
was precisely of the same form as in Eq.~\ref{uc1}.}.
\begin{equation}
U_C=W_C\equiv{\rm Tr}{\cal P}\exp \Big\{ig\int_C dl^iA^i\Big\}
\end{equation}

Taking the contour $C$ to run at infinity, we see that in 
gluodynamics the generator of the magnetic 
$Z_2$ symmetry is the fundamental spatial Wilson loop along the boundary of 
the spatial plane\footnote{There is a slight subtlety here that may be
worth mentioning.
The generator of a unitary transformation should be a unitary
operator. 
The trace of the 
fundamental Wilson loop on the other hand is not unitary. One should therefore
strictly speaking consider instead a
unitarized Wilson loop $\tilde W={W\over \sqrt{WW^\dagger}}$. However 
the factor
between the two operators $\sqrt {WW^\dagger}$ is an operator that is 
only sensitive
to behavior of the fields at infinity. It commutes with all physical 
local operators
$O(x)$ unless $x\rightarrow\infty$. 
In this it is very different from the Wilson loop itself,
which has a nontrivial commutator with vortex operators $V(x)$ at all 
values of $x$.
Since the correlators of all gauge invariant local fields in the pure 
Yang Mills theory
are massive and therefore short range, the operator $\sqrt{W(C)W^\dagger(C)}$ 
at $C\rightarrow\infty$ must be a constant
operator on all finite energy states. The difference between $W$ and 
$\tilde W$ is 
therefore a trivial constant factor and we will not bother with it in the
following.}.

Of course one does not have to go through the exercise with the Georgi
Glashow model in order to show that the fundamental Wilson loop generates
a symmetry. Instead one can directly consider the commutator
\begin{equation}
[W,H]=\lim_{C\rightarrow\infty}\oint_C dx_i
{\rm Tr}{\cal P} E_i(x)\exp\{i\oint_{C(x,x)} dy_iA^i(y)\}
\rightarrow_{C\rightarrow\infty}0
\label{commutator}
\end{equation}
Here the integral in the exponential on the right hand side 
starts and ends at the point of insertion of the electric field. 
For a finite contour $C$ the commutator does not vanish 
only along the contour itself, but it does not contain any  bulk
terms. Making the contour $C$ go to infinity and assuming as usual that 
in a theory with finite mass gap at
infinity no physical modes are excited we conclude that the commutator of
$H$ with infinitely large Wilson loop vanishes\footnote{Note that the
nonvanishing of the commutator at finite $C$ is precisely of the same
nature as for any other
 "conserved charge"
which is defined as an integral of local charge density 
$Q=\lim_{C\rightarrow\infty} 
\int_{|x|\in C}d^2x\rho(x)$. The commutator of such a charge with 
a Hamiltonian also
contains surface terms, since the charge density $\rho$ itself never
commutes 
with the 
Hamiltonian. The commutator is rather a total derivative. 
For a conserved charge, due to the continuity equation
this surface term is equal to the circulation of the spatial component
of 
the current
$$
[Q,H]=\rightarrow_{C\rightarrow\infty}\oint dx^ij_i
$$.
The vanishing of this term again is the consequence of the vanishing
of the 
physical fields
at infinity in a theory with a mass gap. When the charge is not
conserved, 
the commutator
in addition to the surface term contains also a bulk term. It is the
absence 
of such 
bulk terms that is the unique property of a conserved charge. 
The same conclusion is reached if rather than considering the 
generator
of the algebra, one considers the commutator of the group element for 
either continuous or 
discreet symmetry groups.
The commutator in Eq.~\ref{commutator} therefore indeed tells 
us that $W$
is a conserved operator.}.

Next we consider the vortex operator Eq.~\ref{VQCD}. Again 
in order to find the pure Yang-Mills limit of it
we have to integrate 
this expression over 
the orientations of the unit vector $n^a$. 
This integration in fact is equivalent to averaging over the gauge group.
Following ~\cite{dp} one can write  $n_a$ in terms of the SU(2)
gauge transformation matrix $\Omega$.
\begin{equation}
\vec n={1\over 2}{\rm Tr}\Omega\tau\Omega^{\dagger}\tau_3
\label{ucl3}
\end{equation}
The vortex operator in the pure gluodynamics limit then becomes
\begin{equation}
\tilde V(x)=\int D\Omega\exp{{2\pi i\over g}\int_C
dy_i\epsilon_{ij}{\rm Tr}\Omega E_j\Omega^{\dagger}\tau_3}
\end{equation}
This form makes it explicit that $\tilde V(x)$ is defined as the gauge singlet part
of the following, apparently non gauge invariant operator
\begin{equation}
V(x)=\exp {\frac{2\pi i}{g}}\int_C dy^i
\epsilon_{ij}E^3_i(y)
\label{v33}
\end{equation}
The integration over $\Omega$ obviously projects out the gauge singlet part 
of $V$.
In the present case however this projection is redundant. This is because
even though $V$ itself is not gauge invariant,
when acting on 
a physical state it transforms it into another physical 
state\footnote{This is not a trivial statement, since a generic 
non gauge invariant operator has 
nonvanishing matrix elements between the physical and an unphysical sectors.}. 
By physical states
we mean the states which 
satisfy the Gauss' constraint in the pure Yang-Mills theory.
This property of $V$ was noticed by 't Hooft ~\cite{thooft}.
To show this let us consider $V(x)$ as defined 
in Eq.~\ref{v33}
and its gauge transform $V_\Omega=\Omega^\dagger V \Omega$ where
$\Omega$ is 
an arbitrary
nonsingular gauge transformation operator.
The wave functional of any physical state depends only on gauge
invariant 
characteristics
of the vector potential, i.e. only on the values of Wilson loops over
all 
possible 
contours. 
\begin{equation}
\Psi[A_i]=\Psi[\{W(C)\}]
\end{equation}
Acting on this state by the operators $V$ and $V_\Omega$ respectively we obtain
\begin{eqnarray}
V|\Psi>&=&\Psi_V[A_i]=\Psi[\{VW(C)V^\dagger\}] \nonumber \\
V_\Omega|\Psi>&=&\Psi_V^\Omega[A_i]=\Psi[\{V_\Omega W(C)V^\dagger_\Omega\}]
\end{eqnarray}
It is however easy to see that the action of $V(x)$ and $V_\Omega(x)$ on 
the Wilson loop is 
identical - they both multiply it by the phase belonging to the center
of the group
if $x$ is inside $C$ and do 
nothing otherwise.
Therefore
\begin{equation}
V|\Psi>=V_\Omega|\Psi>
\label{prev}
\end{equation} 
for any physical state $\Psi$.
Thus we have
\begin{equation}
\Omega V|\Psi>= \Omega V\Omega^\dagger|\Psi>=
V|\Psi>
\end{equation}
where the first equality follows from the fact that a physical state is invariant under
action of any gauge transformation $\Omega$ and the second equality 
follows from Eq.~\ref{prev}.
But this equation is nothing but the statement that the state
$V|\Psi>$ is physical, i.e. invariant under any nonsingular gauge
transformation. 

We have therefore proved that 
when acting on a physical state the vortex operator creates 
another physical state. 
For an operator of this type the gauge invariant projection only affects
its matrix elements between unphysical states. Since we are only
interested
in calculating correlators of $V$ between physical states, the
gauge projection is redundant and we can freely use $V$ rather than $\tilde V$ to represent the vortex operator.

The formulae of this section can be straightforwardly generalized 
to $SU(N)$ gauge theories. Once again one can start with the Georgi-Glashow like model, 
where  the $SU(N)$ is Higgsed to $U(1)^{(N-1)}$\footnote{In $SU(N)$ 
theories with $N>2$
there in principle 
can be phases separated from each other
due to spontaneous breaking of some global symmetries.
For instance the $SU(3)$ gauge theory with adjoint matter has
a phase with spontaneously broken
charge conjugation invariance
~\cite{bronoff}. Still even in this phase
the confining properties  are the same as in 
the strongly coupled pure Yang-Mills theory,
with the Wilson loop having an area law.}. The construction of the vortex
operator and the generator of $Z_N$ in this case is very similar and the details are
given in ~\cite{kovner1}.
Taking the mass of the Higgs field to infinity again projects the generator onto the 
trace of the fundamental Wilson loop. The vortex operator can be taken as 
\begin{equation}
V(x)=\exp\{{ 4\pi i\over gN} \int_C dy^i
\epsilon_{ij}{\rm Tr}(YE_i(y))
\label{v2}
\end{equation}
where the hyper charge generator $Y$ is defined as
\begin{equation}
Y={\rm diag} \left(1,1,...,-(N-1)\right)
\end{equation}
and the electric field is taken in the matrix notation $E_i=\lambda^aE^a_i$ with
$\lambda^a$ - the $SU(N)$ generator matrices in the fundamental representation.

To summarize this section, 
we have established two important facts. First, 
$SU(N)$ gauge theories in 2+1 dimensions have global $Z_N$ magnetic
symmetry. The generator of this magnetic symmetry group is the fundamental
Wilson loop around the spatial boundary of the system. Second, this
symmetry has a
local order parameter. This
order parameter is a local gauge invariant scalar field which creates
a magnetic vortex of fundamental flux. 

The next question we should ask is whether this global symmetry is at all
relevant for low energy dynamics. 
In the next section we will show that this is indeed the case. We will
calculate the expectation
value of $V$ in confining and nonconfining situations and will
show that confinement is rigidly related 
to the spontaneous breaking of the magnetic symmetry.

\section{The vacuum realization of the magnetic symmetry, the effective
Lagrangian and confinement. Abelian theory.}

We again start our discussion with Abelian theory.
Consider a $U(1)$ gauge theory with scalar matter field
\begin{equation}
{\cal L} =  - {1\over 4}F^2 
+|D_{\mu} \phi |^2 - M^2 |\phi|^2
- \lambda(\phi^*\phi)^2
\label{lscalar}
\end{equation}
Depending on the values of the coupling constant this theory can be
either 
in the 
Coulomb phase with massless photon and logarithmically confined
charges, 
or in the 
Higgs phase which is massive with screened electric charges. 

\subsection{Realizations of the magnetic symmetry.}

Let us start by calculating $<V>$ in the Coulomb phase.
This can be done using the
standard weak coupling perturbation theory.
The expectation value of $V$ is given by the following expression
\begin{equation}
<V(x)>=N^{-1}\lim_{T\rightarrow\infty}\int dA_0<0|e^{iTH}
e^{{\frac{2\pi i}{g}}\int_C dy^i
\epsilon_{ij}E_i(y)}e^{i\int A_0[\partial_iE_i-J_0]} e^{iTH}|0>
\end{equation}
Here $N^{-1}$ is the normalization factor - the usual vacuum-to-vacuum

amplitude,
$|0>$ is the perturbative Fock vacuum and the integral over $A_0$ is the 
standard representation of the projection operator 
which projects the Fock vacuum $|0>$ onto the gauge invariant subspace
which satisfies the Gauss' law. As usual, discretising time, 
introducing resolution of identity
at every time slice and integrating over $E_i$ in the phase 
space path integral this expression can be rewritten as path integral in the 
field space. The result is easy to understand - it is almost 
the same as for the vacuum-to-vacuum amplitude, except that at time $t=0$ the spatial
derivative of the scalar potential $A_0$ is shifted by the $c$-number field 
due to the presence of the vortex operator: 
\begin{equation}
<V(x)>=N^{-1}\int {\cal D}A_\mu\exp \left[-{1 \over 4}\int d^3y [\tilde F_{\mu}(y)-\tilde f_{\mu}(y-x)]^2
+L_{Higgs}
\right]
\label{veuc}
\end{equation}
The $c$-number field $\tilde f_{\mu}$ is the magnetic field of an infinitely thin
magnetic vortex which terminates at point $x$. One can view it as the 
Dirac string of a (three dimensional Euclidean) magnetic monopole. 
\begin{equation}
\tilde f_0 = \tilde f_2 = 0,\   \  
\tilde f_1(y) = {2\pi\over g}\theta(y_1) \delta(y_2) \delta(y_3)
\end{equation}
Thus at weak coupling we have to find the solution of the classical
equations of motion following from the action with the external source
Eq.~\ref{veuc}. The nature of this solution is clear: it is just
a Dirac monopole. The action of this solution is IR finite, since the
contribution of the Dirac string (which normally would be linearly IR
divergent) is canceled by the external source.
\be
<V>=\exp\{-S_{cl}\}
\ee
with
\be
S_{cl}={\Lambda\over g^2}
\ee
Here $\Lambda$ is the ultraviolet cutoff which has to be introduced
since the action of a pointlike monopole diverges in the
ultraviolet. This ultraviolet divergence is benign since it can be
eliminated by the multiplicative renormalization of the vortex
operator~\cite{polchinski}. The important point is that since there is
no divergence in the infrared, the expectation value of $V$ is nonvanishing.
Thus we conclude that in the Coulomb phase of QED the magnetic
symmetry is spontaneously broken.

The spontaneous breaking of a continuous symmetry must be accompanied by
appearance of a massless Goldstone particle. Indeed in QED such a
particle 
exists
- it is the massless photon. The matrix element of the magnetic current 
between the vacuum and the one photon state is
\be
<0|\tilde F_\mu|k_i>=Z^{1/2}(0)k_\mu
\ee
where $Z(0)$ is the on shell photon wave function renormalization. 
This is the standard form of a matrix element of a spontaneously broken
current, with $Z(0)$ playing the role of $f^2_\pi$.

Let us now perform the same calculation in the Higgs phase. The 
path integral representation Eq.~\ref{veuc} is still valid. However
the classical solution that dominates this path integral is now very different.
Since in the Higgs phase the photon has nonzero
 mass $\mu$ the classical action of the
three dimensional monopole in the superconducting medium is 
linearly divergent in the infrared ~\cite{vortex}.  
Essentially the magnetic flux
that emanates from the monopole can not spread out in space(time) but
rather is concentrated inside a flux tube of the thickness $1/\mu$
that
starts at the location of the monopole and goes to the spatial
boundary at infinity. The action of such a field configuration is
proportional to the linear size of the system and diverges in the
thermodynamic limit.
As a result the expectation value of $V$ in the Higgs phase vanishes.
\be
<V>=e^{-L}\rightarrow_{L \rightarrow\infty}0
\ee 
The vortex field correlator similarly 
is given in terms of the classical energy of
a monopole-antimonopole pair in the superconductor. 
The Euclidean
action of this configuration is proportional to the distance between the
monopole and the antimonopole and so
the correlator of the $V$ decays exponentially
\begin{equation}
<V^* (x)V (y) > \sim e^{-M_V|x-y|}
\end{equation}
with $M_V\propto 1/g^2$ being the mass of the Nielsen-Olesen vortex.

This simple calculation can be improved perturbatively.
In the next to leading order
one has to calculate the determinant of the Schr\"odinger operator of a 
particle in the field of a monopole and this corrects the value of the
mass $M_V$. Higher orders in perturbation theory can be calculated, 
 but we will not pursue this calculation here.

The main lesson is that the expectation value of the order parameter
vanishes in the Higgs phase, and thus the magnetic symmetry is unbroken.

\subsection{The low energy effective Lagrangian and 
the logarithmic confinement of electric charges.}

Thus we see that the Coulomb - Higgs phase transition can be described
as 
due to 
restoration of the magnetic $U(1)$ symmetry, the pertinent local order

parameter 
being the vortex operator $V$. 
It should be true then that
the low energy dynamics in the vicinity of the phase transition
is described by the effective low energy Lagrangian. For the $U(1)$ symmetry 
breaking such
an effective Lagrangian can be immediately written down
\begin{equation}
{\cal L}=\partial_\mu V^*\partial^\mu V -\lambda(V^*V-\mu^2)^2 
\label{ldual}
\end{equation}

Although this Lagrangian may seem a bit unfamiliar in the 
context of QED, a little thought convinces one 
that it
indeed describes all relevant light degrees of freedom of the theory.
In the Coulomb phase, where $<V>=\mu\ne 0$, the physical 
particles are interpolated 
by the phase and the radial part of $V$
\be
V(x)=\rho e^{i\chi}
\ee
The phase $\chi$ is of course the 
massless Goldstone boson field, i.e. the photon. The 
fluctuation of the radial component $\rho-\mu$, is the lightest scalar particle,
which in this case is the lightest meson, or scalar positronium.
In the Higgs phase, the field $V$ itself interpolates physical excitations -
the ANO vortices. Of course far from the phase transition the vortices are heavy
and there are other, lighter excitations in the spectrum. The validity of this 
effective
Lagrangian on the Higgs phase side is therefore limited to the narrow
critical region where vortices are indeed the lightest particles.

One type of objects that have not appeared in our discussion so far
are charged particles. Indeed in 2+1 dimensions electrical charges are 
confined, and
therefore we do not expect the charged fields to appear as basic
degrees of freedom in 
the effective low energy Lagrangian. However our original purpose was
precisely to understand the mechanism of confinement through studying the 
effective Lagrangian. Thus if we are unable to identify the charged objects 
in this framework our program is doomed to failure.
Fortunately it is not difficult to understand how charged states are 
represented in the Lagrangian Eq.~\ref{ldual}. The easiest way to do this
is to identify the electric charge through the Maxwell equation
\be
J_\mu={1\over 4}\epsilon_{\mu\nu\lambda}\partial_\nu\tilde F_\lambda
\ee
The dual field strength $\tilde F_\mu$ is 
obviously proportional to the conserved $U(1)$ current
\be
\tilde F_\mu={2\pi\over g}i(V^*\partial V-h.c.)
\ee
The proportionality constant in this relation is dictated by the fact that
in the Higgs phase the magnetic vortices that carry one unit of the $U(1)$
charge, carry the magnetic flux of $2\pi/g$. 
The above two relations give
\be
{\frac{g}{\pi }}J_{\mu }=i\epsilon _{\mu \nu \lambda }\partial _{\nu
}(V^{\ast }\partial _{\lambda }V)
\label{j}
\ee
To calculate the electric charge we integrate the zeroth component of
the current over the two dimensional plane
\be
{g\over\pi}Q=\mu^2\oint_{C\rightarrow\infty} dx_i\partial_i\chi
\label{q}
\ee
The electric charge is therefore proportional to the winding number of
the phase of the field $V$. 

So the charged states do appear in the low energy description in a very natural way.
A charged state is a soliton of $V$ with a nonzero winding number.

\begin{figure}
\hskip 4cm
\psfig{figure=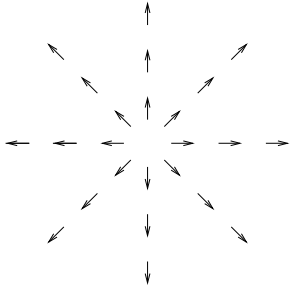,height=1.5in}
\caption{The hedgehog configuration of the field $V$
in the state of unit charge in an Abelian theory.}
\label{fig:hedgehog}
\end{figure}

This identification
tells us immediately that the charged particles are
logarithmically confined. Consider for example the 
minimal energy configuration
in the sector with the unit winding number.
This is a rotationally invariant
hedgehog, Fig.~\ref{fig:hedgehog},
which far from the soliton core has the form 
\begin{equation}
V(x)=\mu e^{i\theta (x)}.
\end{equation}
Here $\theta (x)$ is an angle between the vector $x$  and one of the
axes. 

The self energy of this configuration is logarithmically divergent in
the infrared due to the contribution of the kinetic term 
\begin{equation}
E=\pi \mu ^{2}\ln (\lambda\mu^2L^2)
\label{ech}
\end{equation}
This is nothing but the electromagnetic self energy
of an electrically charged state associated with the logarithmic
Coulomb potential in two spatial dimensions.
The logarithmic confinement of QED is therefore indeed very easily and
transparently seen on the level of the low energy effective Lagrangian.
This in itself perhaps is not such big a deal, since confinement
in this model 
is a linear phenomenon: it is the direct consequence of the logarithmic
behavior of Coulomb potential. We will see later however that this low
energy picture generalizes naturally also
to non Abelian theories and easily accommodates linear confinement.

But before moving on to non Abelian theories, let me make two comments. First,
that in perturbative regime the couplings of the effective Lagrangian can be 
determined in terms of the couplings of the fundamental QED Lagrangian.
To determine the two constants in Eq.~\ref{ldual} one needs two matching 
conditions. One of them can be
naturally taken as the coefficient of the infrared logarithm in the 
self energy
of a charged state.
Matching Eq.~\ref{ech} with the Coulomb self energy gives
\be  
\mu^2={\frac{g^2}{8\pi^2}}  
\label{couplings1}
\ee
The other coefficient is determined by requiring that the mass of the radial
excitation $\rho$ matches the mass of the scalar positronium, which to 
leading order in $g$ is just $2M$.
This condition gives
\be
\lambda={2\pi^2M\over g^2}
\label{couplings2}
\ee
The second comment is about the relation between the vortex operator as 
defined in Eq.~\ref{VQCD} 
and the field $V$ that enters the effective Lagrangian
Eq.~\ref{ldual}.

The vortex operator as defined in Eq.~\ref{VQCD} has a fixed length whereas
the field $V$ which enters the Lagrangian Eq.~\ref{ldual} is a
conventional complex field. How should one understand that? First of
all at weak gauge coupling the quartic coupling in the dual Lagrangian
is large $\lambda \rightarrow \infty $. This condition
freezes the radius of $V$ dynamically. In fact even at finite value of

$\lambda $ if one is interested in the low energy physics, the radial
component is irrelevant as long as it is much heavier than the phase. Indeed
at weak gauge 
coupling the phase of $V$ which interpolates the photon is 
much lighter than all the other excitations in the theory. Effectively
therefore at low energies Eq.~\ref{ldual} reduces to a nonlinear sigma
model and one can identify the field $V$ entering Eq.~\ref{ldual}
directly with the vortex operator of Eq.~\ref{VQCD}. However it is well
known that quantum mechanically the radial degree of freedom of a sigma
model field is always resurrected. The spectrum of such a theory always
contains a scalar particle which can be combined with the phase into a
variable length complex field. The question is only quantitative - how heavy
is this scalar field relative to the phase.

Another way of expressing this is the following. The fixed length field $V$
is defined at the scale of the UV cutoff in the original theory. To arrive
at the low energy effective Lagrangian one has to integrate over all quantum
fluctuations down to some much lower energy scale. In the process of this
integrating out the field is ``renormalized'' and it acquires a dynamical
radial part. The mass of this radial part then is just equal to the mass of
the lowest particle with the same quantum numbers in the original theory.
This is in fact why the parameters in Eq.~\ref{ldual} must be  such that the
mass of the radial part of $V$ is equal to the mass of the lightest
scalar positronium.

\section{Non Abelian theories.}
In this section I want to extend the construction of low energy 
effective Lagrangian to non Abelian theories and see how the realization
of magnetic symmetry in this Lagrangian is related to confinement. Before 
discussing in detail specific models let me present a general argument
that establishes that in a non Abelian theory spontaneous
breaking of magnetic $Z_N$ implies the
area law behavior of the fundamental Wilson loop.

\subsection{Broken $Z_N$ means confinement.}

As we have discussed in the previous section, the generator of magnetic $Z_N$
in pure gluodynamics 
is the fundamental Wilson loop around the spatial boundary of the system.
By the same token the Wilson loop around a closed spatial 
contour $C$ generates the 
$Z_N$ transformation at points inside the contour $C$.
Let us imagine that the $Z_N$ symmetry is spontaneously broken in the 
vacuum and
consider in such a state $|0>$ the expectation value of $W(C)$.
The expectation value $<0|W(C)|0>$ is nothing but the 
overlap of the vacuum state $<0|$
and the state $|S>$ which is obtained by acting with $W(C)$ on the vacuum:
 $|S>=W(C)|0>$. 
If the symmetry is broken, the wave function $|0>$ depends explicitly
on the degrees of freedom which are non invariant under the symmetry
transformation, and is peaked around some specific orientation of these
variables in the group space. For simplicity let us think about all
these non invariant variables as being represented by the vortex field $V$.
The field 
$V$ in the vacuum state has nonvanishing VEV and 
is pointing in some fixed direction in the
internal space. In the state $|S>$ on the other hand its direction in
the internal space is different - rotated by $2\pi/N$ - at points
inside the area $S$ bounded by $C$, since at these points 
the field $V$ has been rotated by 
the action of $W(C)$.
In the local theory with finite correlation length the overlap
between the two states approximately factorizes into the product of the
overlaps taken over the regions of space of linear dimension of order
of the correlation length $l$
\begin{equation}
<0|S>=\Pi_x<0_x|S_x>
\label{fact}
\end{equation}
where the label $x$
is the coordinate of the
point in the center of a given small region of space. For $x$
outside the area $S$ the two states $|0_x>$ and $|S_x>$ are identical and
therefore the overlap is unity. However for $x$ inside $S$ the states
are different and the overlap is therefore some number $e^{-\gamma}$ 
smaller than unity. The number of such regions inside the area is obviously 
of order $S/l^2$
and thus
\begin{equation}
<W(C)>=\exp\{-\gamma{S\over l^2}\}
\end{equation}
In the broken phase the spatial Wilson loop therefore 
has an area law behavior.

Now consider the unbroken phase. Again the average of $W(C)$
has the form of the overlap
of two states which factorizes as in Eq.~\ref{fact}. Now however 
all observables non invariant under $Z_N$  vanish in the vacuum. The action of the 
symmetry generator does not affect the state $|0>$. The state $|S>$ is therefore 
locally exactly the same as the state $|0>$ except along the 
boundary $C$.
Therefore the only regions of space which contribute to the overlap are those
which lay within
one correlation length from the boundary. Thus
\begin{equation}
<W(C)>=\exp\{-\gamma P(C)\}
\end{equation}
where $P(C)$ is the perimeter of the curve $C$.

The crucial requirement for this argument to hold is the existence of a mass
gap in the theory. If the theory contains massless excitations the 
factorization 
of the overlap does not hold, and so in principle even in the broken phase
the Wilson loop can have perimeter behavior. This indeed is the case in
the Abelian theories.

We now turn to the discussion of low energy effective theories.
This will allow us to see the explicit realization of this general argument.

\subsection{The Georgi-Glashow model.}
Let us again start with the Georgi-Glashow model.
For simplicity all explicit calculations in this section will be performed for
the $SU(2)$ gauge theory. Generalization to the $SU(N)$ group is not difficult and
is discussed in ~\cite{kovner1}.

Not much has to be done here to parallel the 
calculation of $<V>$ of the previous section.
The theory is weakly interacting, and all calculations are explicit. Choosing
unitary gauge $n^a=\delta^{a3}$ the perturbative calculation becomes 
essentially
identical to that in the Coulomb phase of QED. The only difference is that 
the charged matter
fields are vectors ($W^\pm_\mu$) rather than scalars ($\phi$), but this only 
enters at the level of the loop corrections.
The nonperturbative monopole contributions are there, but they affect
the 
value
of $<V>$ very little, since $<V>\ne 0$ already in perturbation theory.
Thus just like in the Coulomb phase of QED, $<V>\ne 0$ and the
magnetic 
symmetry 
is spontaneously broken.
The real difference comes only when we ask what is the effective Lagrangian
that describes the low energy physics. Here the monopole contributions are 
crucial, 
since as we have seen before the $U(1)$ magnetic symmetry of QED is explicitly 
(anomalously)
broken by these contributions to $Z_2$. The effective Lagrangian therefore
must have an extra terms which reduce the symmetry of the Eq.~\ref{ldual}.
The relevant effective Lagrangian is
\begin{equation}
{\cal L}=\partial_\mu V^*\partial^\mu V -\lambda(V^*V-\mu^2)^2 
+\zeta(V^2+(V^*)^2)
\label{ldualgg}
\end{equation}

The addition of this extra symmetry breaking 
term has an immediate effect on the mass of the 
"would be" photon - the phase of $V$. Expanding around $<V>=\mu$ we see that
the phase field now has a mass $m_{ph}^2=4\zeta$. This is 
consistent with the classical analysis by Polyakov ~\cite{polyakov}- the monopole
contributions turn the massless photon of QED into a massive (pseudo)scalar
with the exponentially small mass $m_{ph}\propto\exp\{-M_W/g^2\}$.
As a matter of fact for very weak coupling, when the modulus of 
$V$ can be considered as frozen, the Lagrangian Eq.~\ref{ldualgg} in
terms of the phase $\chi$ reduces to Polyakov's dual Lagrangian. The exact 
correspondence between the two is discussed in ~\cite{kovner1}.

\begin{figure}
\hskip 3.3cm
\psfig{figure=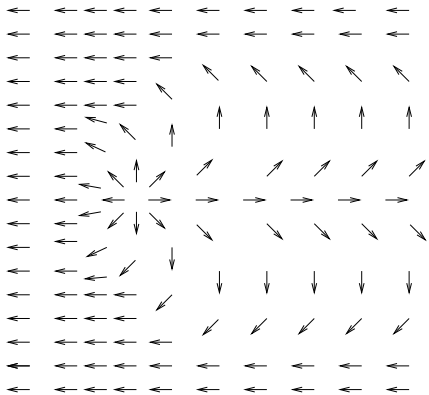,height=2in}
\caption{The  string like configuration of the field $V$
in the state of unit charge in the presence of 
the symmetry breaking terms in the
effective Lagrangian.}
\label{fig:string}
\end{figure}

The explicit symmetry breaking causes a dramatic change in the
topologically charged (soliton) sector. 
We know from Polyakov's analysis that the charges in this model
are confined by linear potential.
This as opposed to QED where confinement is logarithmic. In the 
effective Lagrangian description this is due to the explicit symmetry 
breaking term. The crucial point is that the vacuum of the theory is not
infinitely degenerate $<V>=e^{i\chi}\mu$ with arbitrary constant $\chi$, 
as in the case of QED but only doubly degenerate
$<V>=\pm\mu$. Thus the lowest energy state in the nontrivial winding sector
can not be a hedgehog. In a hedgehog configuration the field $V$ 
at each point in spatial infinity points in a different direction 
in the internal space.
This is OK if all these directions are minima of the potential.
Then the total energy of the configuration comes from the kinetic term, and as
we saw is logarithmic.
However now the potential has only two minima. 
Thus the hedgehog field is
far from the vacuum everywhere in space. The energy of such a state therefore
diverges as the volume of the system: $E\propto g^{2}m^{2}L^{2}$.
Clearly to minimize the energy in a state with 
a nonzero winding, the system must be as in one of the two vacuum states
in as large a region of space as possible. However since the field has to wind
when one goes around the position of a soliton even at arbitrarily large 
distance,
$V$ can not be aligned with the vacuum everywhere at infinity.
The best bet for a system is therefore to choose
a string like configuration Fig.~\ref{fig:string}. The phase of $V(x)$ deviates
from $0$ (or $\pi $) only inside a strip of width $d\sim 1/m$ stretching
from the location of the soliton (charge) to infinity. The energy of such a
configuration diverges only linearly with the dimension $L$. 
In fact a back-of-the-envelope estimate with the effective 
Lagrangian Eq.~\ref{ldualgg} gives the energy
of such a confining string as $E\propto g^2m_{ph}L$.
Clearly the energy  of a soliton and an anti soliton separated
by a large distance $R$ is $E=\sigma R$ with the string tension
$\sigma\propto g^2m_{ph}$. 

This is the simple picture of
confinement in the effective Lagrangian approach in the weakly coupled
regime.

The preceding discussion pertains to confinement of "adjoint" 
color charges. So far we have been considering topological solitons
with unit winding, which corresponds to the charge of the massive $W^\pm$
bosons, or "massive gluons". It should be noted that 
the notion of adjoint string tension is not an absolute one.
Our discussion so far neglected the fact that the solitons have a finite
core energy and therefore in principle can be created in pairs from the vacuum.
Thus the soliton - anti soliton interaction at distance $R$ can be screened 
by creating such a 
pair, if $R$ is big enough. The distance at which the string breaking occurs
can be estimated from the energy balance between the energy stored in the
string $E_S=\sigma_{Adj}R$ and the core energy of the soliton-anti soliton pair,
which in our model is twice the mass of the $W$-boson $2M_W$.
\be
g^2m_{ph}R=2M_W
\ee
The distance at which the string breaks is therefore
\be
R_{breaking}\propto{M_W\over g^2}{1\over M_{ph}}
\ee
Since the width of the string is of order $1/m_{ph}$ and in the weak coupling
$M_W\gg g^2$, the length of the string is indeed much greater than its width.
One can therefore sensibly talk about a
well formed adjoint confining string.

As opposed to adjoint string tension, the concept of the 
fundamental string tension is sharply defined even in principle.
This is  because the theory does not contain
particles with fundamental charge and thus an external fundamental charge can not 
be screened. 
To discuss confinement of external fundamental 
charges we have to learn how to deal with half integer windings. 
Imagine adding to the Georgi-Glashow model some extra 
very heavy fields in the fundamental representation.
The quanta of these fields will carry half integer "electric" charge 
Eq.~\ref{q} and will be confined with a different string tension than $W^\pm$.
To calculate this string tension we should 
consider the Abelian Wilson loop with half integer charge.
We will now do it in the effective theory framework.

Let us first consider a space like Wilson loop. 
As discussed in the previous section, this operator is closely related to
the generator of the magnetic $Z_2$. In fact $W(C)$ 
is nothing but the operator that performs the $Z_2$ transformation inside the
area bounded by the contour $C$.

It is straightforward to write down an operator in the effective theory
in terms of the field $V$
that has the same property 
\begin{equation}
W(C)=e^{i\pi \int_{S}d^{2}xP(x)}  \label{wld}
\end{equation}
Here $S$ is the surface bounded by the contour $C$, and $P$ is the operator
of momentum conjugate to the phase of $V$. In terms of the radius and the phase
of $V$ the path integral representation for calculating the vacuum
average of this Wilson loop is 
\begin{equation}
<W(C)>=\int DV\exp \left\{i\int d^{3}x\rho ^{2}(\partial _{\mu }\chi -j_{\mu
}^{S})^{2}+(\partial _{\mu }\rho )^{2}-U(V)\right\}  
\label{wldpi}
\end{equation}
where $U(V)$ is the $Z_{2}$ invariant potential of Eq.~\ref{ldualgg}. The
external current $j_{\mu }^{S}(x)$ does not vanish only at points $x$ which
belong to the surface $S$ and is proportional to the unit normal $n_{\mu }$
to the surface S. Its magnitude is such that when integrated in the
direction of $n$ it is equal to $\pi $. These properties are
conveniently
encoded in the following expression
\begin{equation}
\int_{T}dx_{\mu }j_{\mu }^{S}(x)=\pi n(T,C)  \label{angle}
\end{equation}
Here $T$ is an arbitrary closed contour, and $n(T,C)$ is the linking number
between two closed curves $T$ and $C$\footnote{Note that here we are
dealing
with
the path integral representation, and thus the contour $C$ and the surface $S$
are embedded into a three dimensional Euclidean space. The linking number 
between two
curves is also defined in three dimensions.}. 

This path integral representation follows immediately if we notice that
the conjugate momentum $P$ is $P(x)=2\rho^2(x)\partial_0\chi(x)$ at some 
fixed time $t$. This accounts for the linear in $\partial_\mu\chi$ in the 
exponential Eq.~\ref{wldpi}. The constant term $j^2$ arises due to standard
integration over the conjugate momenta in arriving to the path integral
representation.

The path integral representation was constructed for
the  spatial Wilson loops. However the expression Eq.~\ref{wldpi}
is completely covariant, and in this form is valid for time like
Wilson loops as well. It is important to note that although the expression
for the current depends on the surface $S$, the Wilson loop operator in fact
depends only on the contour $C$ that bounds this surface. A simple way to
see this is to observe that a change of variables
$\chi\rightarrow\chi+\pi$ 
in the
volume bounded by $S+S^{\prime}$ leads to the change $j_\mu^S\rightarrow
j_\mu^{S^{\prime}}$ in eq.(\ref{wldpi}). The potential is not affected by
this change since it is globally $Z_2$ invariant. Therefore the operators
defined with $S$ and $S^{\prime}$ are completely equivalent.

\begin{figure}
\hskip 2.5cm
\psfig{figure=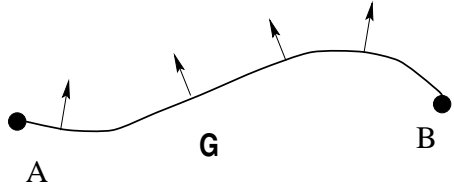,height=1in}
\caption{The external current $j_\mu$
which creates the pair of static fundamental charges 
in the effective Lagrangian description.}
\label{fig:current}
\end{figure}

To calculate the energy of a pair of static fundamental charges at
points $A$ and $B$ we have to consider a time like fundamental Wilson loop of
infinite time dimension. This corresponds to time independent $j_\mu$ which
does not vanish only along a spatial curve $G$ (in equal time cross section)
connecting the two points and
pointing in the direction normal to this curve Fig.\ref{fig:current}. 
The shape of the curve
itself does not matter, since changing the curve without changing its
endpoints is equivalent to changing the surface $S$ in Eq.~\ref{wldpi}.

\begin{figure}
\hskip 2cm
\psfig{figure=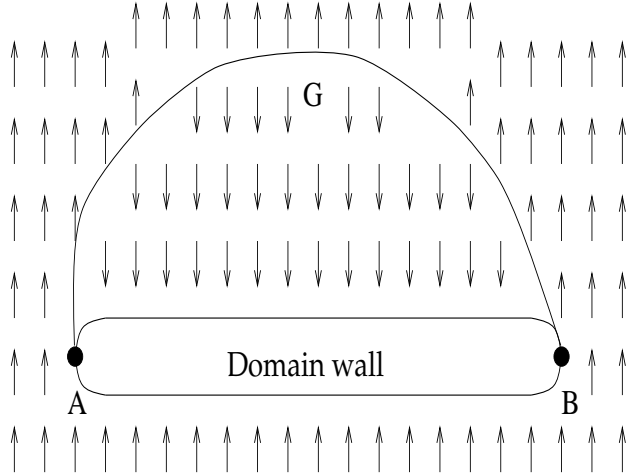,height=2.5in,width=3.2in}
\caption{The minimal energy configuration of $V$ in the presence of a
pair of fundamental charges.}
\label{fig:wall}
\end{figure}

In the classical approximation the path integral Eq.~\ref{wldpi} is
dominated by a static configuration of $V$. To determine it we have to
minimize the energy on static configurations in the presence of the external
current $j_{\mu }$. The qualitative features of the minimal energy solution
are quite clear. The effect of the external current is to flip the
phase of $V$ by $\pi $ across the curve $G$, as is expressed in 
Eq. ~\ref{angle}. Any configuration that does not have this behavior
will 
have the
energy proportional to the length of $G$ and to the UV cutoff scale.
Recall that the vacuum in the theory is doubly degenerate. The
sign change 
of $V$ transforms one vacuum configuration into the other one. The
presence of $j_{\mu }$ therefore requires that on opposite sides of the
curve $G$, immediately adjacent to $G$ there should be different vacuum
states. It is clear however that far away from $G$ in either direction the
field should approach the same vacuum state, otherwise the energy of a
configuration diverges linearly in the infrared. The phase of $V$ therefore
has to make half a wind somewhere in space to return to the same vacuum
state far below $G$ as the one that exists far above $G$. If the distance
between $A$ and $B$ is much larger than the mass of the lightest particle in
the theory, this is achieved by having a segment of a domain wall between
the two vacua connecting the points $A$ and $B$. Clearly to minimize the
energy the domain wall must connect $A$ and $B$ along a straight line.
The energy of such a domain wall is proportional to its length, and
therefore the Wilson loop has an area law behavior. The minimal energy
solution is schematically depicted on Fig.~\ref{fig:wall}.
We see that the string tension for the fundamental string is equal to the
tension of the domain wall which separates the two vacua in the theory. 
This relation has been discussed a long time ago by 't Hooft ~\cite{thooft}.
Parametrically this string tension is clearly the same as the adjoint one,
\be
\sigma_f\propto g^2m_{ph}
\ee
although the proportionality constant is different. We will briefly discuss
the relation between the adjoint and the fundamental string tensions 
in the next subsection.

Note that
the fundamental string is an absolutely stable topological
object in the $Z_{2}$ invariant theory: the domain wall. It can not break,
if one makes the distance between the two charges larger. In the
effective 
theory it is also
obvious since there is no pointlike (particle like)
object in the theory on which a domain wall can
terminate since there are no dynamical objects with half an integer 
winding number.

We interpreted this calculation as the calculation of the potential between
the fundamental adjoint charges - the time like Wilson loop. However in
the Euclidean formulation there is no difference between time like and 
space like
Wilson loops. Interpreted in this way the calculation becomes the technical
illustration to the argument given in the previous subsection: space like Wilson
loop has an area law if the magnetic $Z_2$ is spontaneously broke.

\subsection{Gluodynamics.}

In the weakly interacting case the
effective low energy Lagrangian can be derived as explained
above in perturbation theory plus dilute monopole gas approximation.
The more interesting regime is of course that of strong coupling,
which is essentially the pure Yang-Mills theory.
The luxury one has in 2+1 dimensions is that
the weak and the strong coupling regimes are
not separated by a phase transition~\cite{fradkin}. 
This means that whatever global symmetries the theory has, their
realization must be
the same in the weakly coupled and the strongly coupled vacua.
The existence of the $Z_{N}$ symmetry  is an exact
statement which is not related to the weak coupling limit. It is therefore
natural to expect that this symmetry must be nontrivially
represented in the effective low energy Lagrangian. 
It is plausible then that the low
energy dynamics at strong coupling is described by the same effective
Lagrangian which encodes spontaneous breaking of the magnetic symmetry.
The values of the coupling constants will be of course different in
the two regimes, but the qualitative behavior should be similar.

Strictly speaking this is an
assumption and not a theorem. That is where the difference between 
continuous and
discreet symmetries comes in. If the magnetic symmetry were continuous
(like in the Abelian case) its spontaneous breaking would unambiguously
determine the structure of the effective Lagrangian, be the theory
weakly or strongly coupled.
With discrete symmetry this is not necessarily the case.
 It could
happen that even though the symmetry is broken, the ``pseudo Goldstone"
particle is so heavy that it decouples from the low energy dynamics.
For this to happen though, the symmetry breaking would have to occur on 
a very high energy scale.
In gluodynamics this is very unlikely, since the theory has only one 
dynamical scale. 
In fact as we have seen in the beginning of this section, the fundamental
string tension determines the scale at which $Z_N$ is broken. 
Wilson loops of linear size $l\le(\sigma)^{-1/2}$ do not distinguish between
confining and nonconfining behavior, and thus between the broken and the unbroken
$Z_N$. The scale of $Z_N$ breaking therefore is $(\sigma)^{1/2}$
which is precisely the natural dynamical scale of QCD.

Generically therefore we expect that the 
pseudo Goldstone stays among the low energy excitations not only in weakly 
coupled
limit but also in pure gluodynamics.
If this is the case
the degrees of freedom that enter
the effective Lagrangian in weakly coupled phase also
interpolate real low energy physical states of the
strong coupling regime. That is to say the radial and phase
components of the vortex field $V$ must correspond to lightest glueballs
of pure $SU(N)$ Yang Mills theory. 
We can check whether this is the case by 
considering the lattice gauge theory data on the spectrum~\cite{teper}. 
The radial part of $V$ is
obviously a scalar and has quantum numbers $0^{++}$. The quantum numbers of
the phase are easily determined from the definition Eq.~\ref{VQCD}. Those
are $0^{--}$. The spectrum of pure $SU(N)$ Yang Mills theory in 2+1
dimensions was extensively studied recently on the lattice ~\cite{teper}. The
two lightest glueballs for any $N$ are found to have exactly those quantum
numbers. The lightest excitation is the scalar while the next one is a
charge conjugation odd pseudo scalar with the ratio of the masses roughly $
m_{p}/m_{s}=1.5$ for any $N$\footnote{Actually 
this state of matters is firmly established only for $N>2$. At $N=2$
the mass of the pseudo scalar has not been calculated in ~\cite{teper}. The
reason is that it is not clear how to construct a charge conjugation odd
operator in a pure gauge $SU(2)$ lattice theory. So it is possible that the
situation at $N=2$ is non generic in this respect. In this case 
our strong coupling picture should apply at $N>2$.}.

The situation therefore likely is the following. The low energy physics of
the $SU(2)$ gauge theory is always described by the effective
Lagrangian 
Eq.~\ref{ldualgg}. In the weak coupling regime the parameters are given in 
Eqs.~\ref{couplings1},\ref{couplings2}. Here the pseudo scalar particle 
is the lightest state in the spectrum
and the
scalar is the first excitation. The pseudo scalar
is the almost massless "photon" and the scalar is the massive Higgs
particle. Moving towards the strong coupling regime (decreasing the Higgs
VEV in the language of the Georgi-Glashow model) 
leads to increasing the pseudo scalar
mass while reducing the scalar mass and the parameters of the effective
Lagrangian change accordingly. The crossover between the weak and the strong
coupling regimes occurs roughly where the scalar and the pseudo scalar become
degenerate. At strong coupling the degrees
of freedom in the effective Lagrangian are the two lightest glueballs. They
are however still collected in one complex field which represents
nontrivially the exact $Z_{2}$ symmetry of the theory (or $Z_N$ for
$SU(N)$). 

Of course the spectrum
of pure Yang Mills theory apart from the scalar and the pseudo scalar glueballs
contains many other massive glueball states and those are not separated by a
large gap from the two lowest ones. Application of the effective Lagrangian
in the strong coupling regime therefore has to be taken in a qualitative
sense. On this qualitative level though, as we have seen linear confinement
is an immediate property of this type of effective Lagrangian.

The fact that masses of
the scalar and 
the pseudo scalar particles are interchanged in gluodynamics relative to the
weakly coupled regime leads to some interesting qualitative
differences ~\cite{kovner3}.
In particular the structure of the confining string and the interaction
between the strings differ in some important ways.
Let me briefly discuss those.

\begin{figure}
\hskip 2cm
\psfig{figure=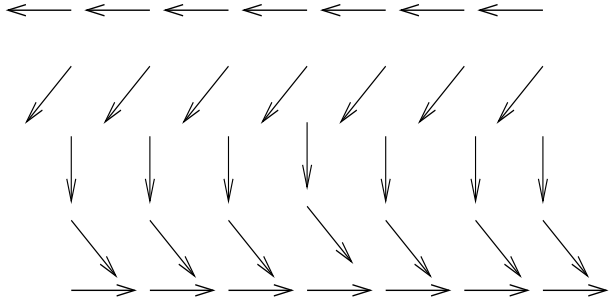,height=1.5in,width=3.2in}
\caption{The structure of the string (domain wall) in the regime when
the pseudo scalar is lighter than the scalar, $m<M$.}
\label{fig:weakwall}
\end{figure}

In the weakly coupled regime the phase of $V$ is
much lighter than the radial part. A cartoon of the fundamental string in
this situation is depicted in Fig.~\ref{fig:weakwall}.
The radial part $\rho$ being very heavy practically does not change inside
the string. The value of $\rho$ in the middle of the string can
be estimated from the following simple argument. The width of region where $
\rho$ varies from its vacuum value $\mu$ to the value $\rho_0$ in the middle
is of the order of the inverse mass of $\rho$. The energy per unit length
that this variation costs is
\begin{equation}
\sigma_\rho\sim M(\mu-\rho_0)^2+x{\frac{m^2}{M}}\rho_0^2
\end{equation}
where $x$ is a dimensionless number of order unity.
The first term is the contribution of the kinetic term of $\rho$ and the second
contribution comes from the interaction term between $\rho$ and $\chi$ due
to the fact that the value of $\chi$ in the middle of the string differs
from its vacuum value. 
Our notations are such that $m$ is the mass of the pseudo scalar 
particle and $M$ is
the mass of the scalar.
Minimizing this with respect to $\rho_0$ we find 
\begin{equation}
\rho_0=\mu (1-x{\frac{m^2}{M^2}})
\end{equation}
Thus even in the middle of the string the difference in the
value of $\rho$ and its VEV is second order in the small ratio $m/M$.
Correspondingly the contribution of the energy density of $\rho$ to the
total energy density is also very small. 
\begin{equation}
\sigma_\rho\sim {\frac{m }{M}}m\mu^2
\end{equation}
This is to be compared with the total tension of the string which is
contributed mainly by the pseudo scalar phase $\chi$ 
\begin{equation}
\sigma_\chi\sim m\mu^2
\end{equation}
This again we obtain by estimating the kinetic energy of $\chi$ on a
configuration of width $1/m$ where $\chi$ changes by an amount of order $1
$\footnote{The fact that the heavy radial field $\rho $ practically does not contribute to
the string tension is natural from the point of view of
decoupling. In the limit of infinite mass $\rho $ should decouple from the
theory without changing its physical properties. It is however very
different from the situation in superconductors. In a superconductor of the
second kind, where the order parameter field is much heavier than the
photon ( $\kappa >
\frac{1}{\sqrt{2}}$) the magnetic field and the order parameter give
contributions of the same order
(up to logarithmic corrections $O(\log \kappa) $)
to the energy of the Abrikosov
vortex. This is the consequence of the fact
that the order parameter itself is forced to vanish in the core of the
vortex, and therefore even though it is heavy, its variation inside
the vortex is large. An even more spectacular situation arises if we
consider a domain wall between two vacuum states in which the heavy field
has different values ~\cite{kovner4}. In this situation the contribution of
the heavy field $\phi$ to the tension would be
\begin{equation}
\sigma _{heavy}=M(\Delta \phi)^{2}
\end{equation}
where $\Delta \phi$ is the difference in the values of $\phi$ 
on both sides of the
wall. For fixed $\Delta \phi$ the energy density diverges when $\phi$ 
becomes
heavy. In the present case this does not happen since the two vacua which are
separated by the domain wall differ only in VEV of the light field $\chi $
and not the heavy field $\rho $.}.

\begin{figure}
\hskip 2.5cm
\psfig{figure=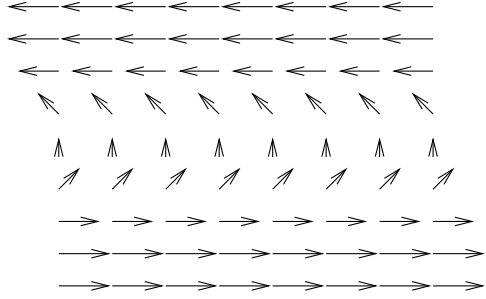,height=1.5in}
\caption{The structure of the string (domain wall) in the regime when
the pseudo scalar is heavier than the scalar, $m>M$.}
\label{fig:strongwall}
\end{figure}

Let us now consider the domain wall (or fundamental string) in the 
opposite regime, that is
when the mass of the scalar is much smaller than the mass of the
pseudo scalar. The profile of the fields in the wall now is very different.
The cartoon of this situation is given on Fig.~\ref{fig:strongwall}
We will use the same notations, denoting the mass of
the pseudo scalar by $m$ and the mass of the scalar by $M$, but now $m\gg M$.
Let us again estimate the string tension and the contributions of the scalar
and a pseudo scalar to it. The width of the region in which the
variation 
of $\rho $ takes place is of the order of its inverse mass. The estimate of the
energy density of the $\rho $ field is given by the contribution of the
kinetic term 
\begin{equation}
\sigma _{\rho }\sim M(\mu -\rho _{0})^{2}
\end{equation}
The width of the region in which the phase $\chi $ varies is $\sim 1/m$. In
this narrow strip the radial field $\rho $ is practically constant and is
equal to $\rho _{0}$. The kinetic energy of $\chi $ therefore contributes 
\begin{equation}
\sigma _{\chi }\sim m\rho _{0}^{2}
\end{equation}
Minimizing the sum of the two contributions with respect to $\rho _{0}$ we
find 
\begin{equation}
\rho _{0}\sim {\frac{M}{m}}\mu \ll\mu 
\end{equation}
And also 
\begin{eqnarray}
\sigma _{\chi } &\sim &{\frac{M}{m}}M\mu ^{2}  \nonumber \\
\sigma _{F} &=&\sigma _{\rho }\sim M\mu ^{2}
\end{eqnarray}

Now the radial field is very small in the core of the string.
The energy density is contributed almost entirely by the scalar rather
than by the
pseudo scalar field\footnote{Again this is in agreement with
decoupling. 
The heavier
field does not contribute to the energy, even though its values
on the opposite
sides of the wall differ by $O(1)$. Its contribution to the energy
is suppressed by the factor $\rho _{0}^{2}$ which is very small inside
the wall.}.

The extreme situation $m\gg M$  is not
realized in the non Abelian gauge theory. From the lattice simulations we
know that in reality even in the pure Yang Mills case the ratio between the
pseudo scalar and scalar masses is about $1.5$ - not a very large number. The
analysis of the previous paragraph therefore does not reflect the situation
in the strongly coupled regime of the theory. Rather we expect that the
actual profile of the string is somewhere in between
Fig.~\ref{fig:weakwall} 
and Fig.~\ref{fig:strongwall}
although somewhat closer to Fig.~\ref{fig:strongwall}. 
The widths of the string in terms of the
scalar and pseudo scalar fields are of the same order, although the scalar
component is somewhat wider. The same goes for the contribution to the
string tension. Both glueballs contribute, with the scalar contribution being
somewhat larger\footnote{One expects similar non negligible contributions
also from higher mass glueballs which are not taken into 
account in our effective Lagrangian framework.}. 

The interaction between two 
domain walls (confining fundamental strings) in the two extreme
regimes is also quite different. In the weakly coupled region we
can disregard the variation of $\rho $. For two widely separated parallel
strings the
interaction energy comes from the kinetic term of $\chi $. This obviously
leads to repulsion, since for both strings in the interaction region the
derivative of the phase is positive. On the other hand if the pseudo scalar
is very heavy the main interaction at large separation is through the
''exchange'' of the scalar. This interaction is clearly attractive, since if
the strings overlap, the region of space where $\rho $ is different from its
value in the vacuum is reduced relative to the situation when the strings
are far apart.

The situation is therefore very similar to that in superconductivity. The
confining strings in the weakly coupled and strongly coupled regimes behave
like Abrikosov vortices in the superconductor of the second and first kind
respectively. This observation has an immediate implication for the string
tension of the adjoint string. As we have discussed in the previous section
the phase $\chi $ changes from $0$ to $2\pi $ inside the adjoint string. The
adjoint string therefore can be pictured as two fundamental strings running
parallel to each other. In the weak coupling regime the two
fundamental 
strings
repel each other. The two fundamental strings within the adjoint string
therefore will not overlap, and the energy of the adjoint string is twice
the energy of the fundamental one. 
\begin{equation}
\sigma _{Adj}=2\sigma _{F}
\end{equation}

In the strongly coupled region the situation is quite different. The strings
attract. It is clear that the contribution of $\rho $ to the energy will be
minimized if the strings overlap completely. In that case the contribution
of $\rho $ in the fundamental and adjoint strings will be roughly the same.
There will still be repulsion between the pseudo scalar cores of the two
fundamental strings, so presumably the core energy will be doubled. We
therefore have an estimate: 
\begin{equation}
\sigma _{Adj}=\sigma _{F}+O({\frac{M}{m}}\sigma _{F}).  \label{sst}
\end{equation}

Again in the pure Yang - Mills limit the situation is more complicated. The
scalar is lighter, therefore the interaction at large distances is
attractive. However the pseudo scalar core size and its contribution to the
tension is not small. In other words $M/m$ in Eq.~\ref{sst} is a number
of order one.

Thus broadly the relation between the weakly and strongly coupled 
confining theories is similar to the relation of the superconductors of 
the second and first kind.

\section{Magnetic $Z_N$ in 3+1 dimensions.}

The discussion in the previous three sections was in the framework of 
2+1 dimensional theories.
The reason for this is that magnetic symmetry in 2+1 dimensions is simple.
The objects that carry the magnetic charge are particles which are interpolated
by local scalar fields, and they are natural building blocks for the 
effective theory.

In 3+1 dimensions the situation is a little more complicated. The objects 
that carry 
magnetic symmetry are strings rather than particles.
Still many elements of the 2+1 dimensional discussion can be generalized.

\subsection{Magnetic symmetry generator and the vortex operator in 3+1
dimensions.}

First of all, 't Hooft's topological argument carries over to 3+1 dimensions 
immediately. However here it establishes the existence of topologically
stable string like objects rather than particles. The gauge
transformation
$U(x)$ winds once in the group space when we go around any closed
curve which
encloses the core of such a fluxon.

The symmetry that guarantees stability of these objects 
is again associated with the
magnetic field.
In Abelian theories the symmetry is continuous and the
relevant conserved current this time is the dual field strength tensor 
\be
\partial_\mu\tilde F_{\mu\nu}=0
\ee
The somewhat unusual property of this current is that it is a tensor 
and not a vector.
Therefore if we want to define a conserved Lorentz scalar
charge associated with it,
we should integrate the "charge density" not over all the three dimensional
space, but rather over a two dimensional space like surface.
The conserved charge then is the magnetic flux through 
such an infinite surface, whose boundary is at infinity
\be
\Phi=\int d^2 S_iB_i
\ee
In an Abelian theory without monopoles this charge does not depend on the 
integration surface as long as the
boundary of the surface remains fixed. 
The symmetry group element as in 2+1 dimensions is large Wilson loop
\be
W_\alpha=\exp\{i\alpha g\oint_{C'\rightarrow\infty} dl_iA_i\}
\ee
The objects that carry this charge are obviously ANO vortices in the 
Higgs (superconducting) phase.

The operator that creates a closed magnetic vortex along a curve $C$
is 
constructed
as
\begin{equation}
V(C)=\exp\{{2\pi i\over g}\int_S d^2 S^i E^i\}
\label{v4}
\end{equation}
where the integration goes over the area $S$ bounded by the contour $C$.
In this definition $g$ is the smallest electric charge present in the theory.
This ensures independence of $V$ on the choice of $S$, as long as its boundary is $C$.
The argument here is exactly the same as the 
one that established independence of $V$ in 2+1 
dimensions of the choice of the cut in the angle function.
With this definition of the 't Hooft loop operator $V(C)$ we can 
easily calculate its 
commutator with the spatial Wilson loop, the so-called 't Hooft 
algebra ~\cite{thooft}
\begin{eqnarray}
V^\dagger(C)W_\alpha(C')V(C)&=&e^{2\pi \alpha i n(C,C')}W_\alpha(C')\nonumber\\
W_\alpha(C')V(C)W^\dagger_\alpha(C')&=&e^{2\pi \alpha i n(C,C')}V(C)
\end{eqnarray}
where $n(C,C')$ is the linking number of the curves $C$ and $C'$.
For an infinite contour $C$ and the Wilson loop along the spatial
boundary of the system
the linking number is always unity.
The $V(C)$ for an infinite loop is therefore an eigenoperator
of the magnetic symmetry and is the analog of the vortex
operator 
$V(x)$
in 2+1 dimensions. Any closed vortex loop of fixed size commutes with the 
Wilson loop if the contour $C'$ is very large. Such a closed loop is thus
an  analog of the 2+1 dimensional vortex-anti vortex correlator
$V(x)V^\dagger(y)$, which also commutes with the global symmetry
generator, 
but has a
nontrivial
commutator with $W_C$ if $C$ encloses only one of the points $x$ or $y$.

Moving on to non Abelian theories it is {\it deja vu} all over again.
The magnetic symmetry is now discreet. A simple way to understand it is again
through monopoles. In an $SU(N)$ theory one finds 't Hooft-Polyakov monopoles 
of magnetic charge ${2\pi N\over g}$. The elementary magnetic fluxon
can 
therefore
disappear in N-plets leaving a monopole-antimonopole pair behind.
This means that the magnetic flux is only preserved modulo $N$, and thus the 
symmetry is reduced to $Z_N$.
The generator of this symmetry is still the large fundamental Wilson loop.
The 't Hooft loop - magnetic vortex creation operator is similar to QED
and 
can be thought
just like in 2+1 dimensions as an operator of a singular gauge 
transformation 
\begin{equation}
V(C)=\exp\{{ i\over gN} \int d^3x {\rm Tr}(D^i\omega_CY) E^i\}=
\exp\{{4\pi i\over gN}\int_S d^2 S^i {\rm Tr} (YE^i)\}
\label{v3}
\end{equation}
with $\omega_C(x)$, the singular gauge function 
which is equal to the solid angle subtended by $C$ as seen from the 
point $x$\footnote{The derivative term $\partial^i\omega$ 
in this expression should be understood to contain only the smooth part of 
the derivative
and to exclude the contribution due to the discontinuity of $\omega$ on
 the surface $S$.}. The function $\omega$ is continuous 
everywhere, except on a surface $S$
bounded by $C$, where it jumps by $4\pi$. Other than the 
fact that $S$ is bounded by 
$C$, its location is arbitrary. 
The vortex loop and the spatial Wilson loop satisfy the 't Hooft algebra
\begin{equation}
V^\dagger(C)W(C')V(C)=e^{{2\pi i\over N} n(C,C')}W(C')
\end{equation}
Again, for an infinite curve $C$ which goes through the whole system
the vortex operator is an eigenoperator of the large Wilson loop and
thus the order parameter of the magnetic $Z_N$ symmetry.

So there is a symmetry and there is an order parameter. Except that
the symmetry charge is not a volume integral but a surface integral,
and the order parameter is not a local field, but a field that creates
a string like object.
This situation is a little unusual, since strictly speaking there are
no phases in such a theory for which the order parameter has a
nonvanishing expectation value in the thermodynamic limit.
This is due to the fact that 
in the system infinite in all directions  $C$ is necessarily 
an infinite line. The expectation value $<V(C)>$
clearly vanishes irrespective of whether $Z_N$ is broken or not.
The 't Hooft loop along a closed contour on the other hand is never
zero, since it is globally invariant under the $Z_N$ transformation. 

Nevertheless the behavior of the closed loops does 
reflect the realization mode of the symmetry, since it is
qualitatively different in the two possible phases. 
Namely vacuum expectation value of a large closed 't Hooft loop 
(by large, as usual we mean  
that the linear dimensions of the loop are much larger than the
correlation length in the theory) has an area law decay if the
magnetic symmetry is spontaneously broken, and perimeter law decay if
the vacuum state is invariant under the magnetic $Z_N$.

To understand the physics of this 
behavior
it is useful to think of the 't Hooft line
as built of ``local'' operators - little ``magnetic dipoles''.
Consider Eq. ~\ref{v3} with the contour $C$ running
along
the x- axis. 
Let us mentally divide the line into (short) segments of length $2\Delta$
centered at $x_i$. Each one of these segments is a little magnetic
dipole and the 't Hooft loop is a product of the operators that
create these dipoles. The definition of these little dipole operators 
is somewhat ambiguous but since we only intend to use them
here
for the purpose of an intuitive argument any reasonable definition will do.
It is convenient to define a single dipole operator in the following way
\begin{equation}
D_\Delta(x)=
\exp
\{i\int d^3y [a_i^+(x+\Delta-y)+a_i^-(x-\Delta-y)]
{\rm Tr} (YE^i(y))\}
\label{d}
\end{equation}
where $a_i^\pm(x-y)$ is the c-number vector potential of the 
Abelian magnetic monopole (antimonopole) of strength $4\pi/gN$. 
The monopole field corresponding to $a_i$ contains both, the 
smooth $x_i/x^3$ part as well as the Dirac string contribution. 
The Dirac string
of the monopole - antimonopole pair in 
Eq. ~\ref{d} is chosen so that is connects the points $x-\Delta$ and
 $x+\Delta$ along the straight line.
The dipole operators obviously have the property
\begin{equation}
D_\Delta(x)
D_\Delta(x+2\Delta)
=D_{2\Delta}(x+\Delta)
\end{equation}
This is because in the product the smooth field contribution of the 
monopole in
$D_\Delta(x)$ cancels the antimonopole contribution in 
$D_\Delta(x+2\Delta)$, while the Dirac string now stretches between 
the points
 $(x-\Delta)$ and $(x+3\Delta)$. When multiplied over the closed contour, 
the smooth fields cancel out completely, while the surviving Dirac string 
is precisely the magnetic vortex created by a closed 't Hooft loop operator. 
The 't Hooft loop can therefore be written as
\begin{equation}
V(C)=\Pi_{x_i}D_\Delta(x_i)
\label{vd}
\end{equation}

The dipole operator $D(x_i)$ is an eigenoperator of the magnetic flux
defined on a surface that crosses the segment $[x_i-\Delta,
x_i+\Delta]$.
 Suppose the magnetic symmetry is broken. Then we expect the dipole operator
to have a nonvanishing expectation value\footnote{The magnetic dipole 
operators defined above are strictly speaking not local, since they carry
the long range magnetic field of a dipole. However, the dipole 
field falls off with distance very fast. Therefore even though this fall off
is not exponential the slight non locality of $D$ should not affect the 
following qualitative discussion.}
 $<D>=d(\Delta)$. If there are no
massless excitations in the theory, the operators $D(x_i)$ and
$D(x_j)$ should be decorrelated if the distance $x_i-x_j$ is greater
than the correlation length $l$. Due to Eq.~\ref{vd}, the 
expectation value of the 't Hooft loop
should therefore roughly behave as
\begin{equation}
<V(C)>=d(l)^{L/l}=\exp\{-\ln\Big({1\over d(l)}\Big) {L\over l}\}
\label{vline}
\end{equation} 
where $L$ is the perimeter of the loop.
In the system of finite length $L_x$, 
the vacuum expectation value of the vortex line which winds around the
system in $x$-direction is therefore finite as in Eq.~\ref{vline}
with $L\rightarrow L_x$.

On the other hand in the unbroken phase 
the VEV of the dipole operator depends on the size of the system in
the perpendicular plane $L_y$. For large $L_y$ it must vanish
exponentially as
$d=\exp\{-a L_y\}$. 
So the expectation
value of $V$  behaves at finite $L_y$ in the unbroken phase as:
\begin{equation}
<V(C)>=\exp\{-a L_yL_x\}
\label{vlinebehave}
\end{equation}
and vanishes as $L_y\rightarrow\infty$.
Thus in a system which is finite in $x$ direction,
but infinite in $y$ direction, the 't Hooft line in the $x$ direction
has a finite VEV in the broken phase and vanishing VEV in the
unbroken phase.

In the limit of the infinite system size $L_x\rightarrow\infty$
the VEV obviously vanishes in both phases. 
This is of course due to the fact that $V$ is a product of infinite
number
of dipole operators, and this product vanishes even if individual
dipole operators have finite VEV\footnote{The VEV of
the dipole $D$ must be smaller than one since $D$ is defined as a
unitary operator.}.
However one can avoid any reference to finite size system
and infinite vortex lines by considering closed 't Hooft loops.

For a closed loop with long sides along $x$ axis at $y=0$ and $y=R$
the above argument leads to the conclusion  
that in the broken phase $V$ must have a
perimeter law, Eq.~\ref{vline}. 
In the unbroken phase the correlation between the
dipoles at $y=0$ and dipoles at $y=R$ should decay exponentially
$<D(0)D(R)>\propto\exp\{-\alpha{R\over l}\}$
and thus 
\begin{equation}
<V(C)>=\exp\{-\alpha {LR\over l^2}\}=\exp\{-\alpha {S \over l^2}\}
\end{equation}
Thus the perimeter behavior of $<V(C)>$ indicates a vacuum state which
breaks spontaneously the magnetic $Z_N$ while the area behavior means
that the magnetic $Z_N$ is unbroken.

The magnetic symmetry structure thus generalizes to 3+1 dimensions.
Moreover 't Hooft ~\cite{thooft} gave an argument according to which
Wilson loop and 't Hooft loop can not simultaneously have an area law
behavior, at least for $N=2,3$. It follows then that in the confining
phase the 't Hooft loop must have a perimeter law behavior, which
signals spontaneous breaking of the magnetic $Z_N$. 
It is reasonable to expect therefore that if we learned how to
implement spontaneous breaking of a symmetry of this type in the
effective Lagrangian framework, we would have a similar universal and
simple picture of confinement as in 2+1 dimensions.
Unfortunately we don't know how to do it at this point. One way of
going about it would be to go in the direction of a string theory,
since the basic operator $V$ creates a string like object. This however
seems unnecessarily complicated for two reasons. First because even if
such a theory can be constructed it would be difficult to get
information out of it. Solving a string theory is not an easy
task, especially since the vacuum of such a theory must be nontrivial
and contain some sort of string  condensate $<V>$. 
And the second reason is that even in
the string theory the low energy dynamics is
always described by some effective field theory. So we may just as
well go directly for an effective field theory description.

Construction of such an effective theory is an interesting open problem.
But even though we can not really do it, we can try to learn
something by taking hints from the lattice QCD. The spectrum of pure
gluodynamics has been quite extensively studied on the lattice. It is
known that the two lightest glueballs are the scalar and the tensor
ones~\cite{glueballs}. We can use this information and try to write
down a simple theory that describes these excitations~\cite{kovner5}. 
The object of
this exercise is to see whether in the framework of such an effective
theory we can identify naturally appearing classical objects
that look like confining strings. Let me briefly describe what happens
if we do that.

\subsection{Effective Lagrangian.}
We start therefore by writing down a theory which contains a scalar 
field $\sigma$ and a massive
symmetric tensor field $G_{\mu\nu}=G_{\nu\mu}$ with a simple interaction. 
\begin{eqnarray}
\label{model}
L&=&{1\over
4}G_{\lambda\sigma}D^{\lambda\sigma\rho\omega}G_{\rho\omega}+\partial_\mu
\sigma\partial^\mu\sigma-2g^2 v\sigma G^{\mu\nu}G_{\mu\nu}\nonumber\\
&-&g^2\sigma^2  G^{\mu\nu}G_{\mu\nu}-V(\sigma)
\end{eqnarray}

The operator $D$ which appears in the kinetic term of the tensor field is
\begin{eqnarray}
\label{D}
D^{\lambda\sigma\rho\omega}&=&(g^{\lambda\rho}g^{\sigma\omega}+
g^{\sigma\rho}g^{\lambda\omega})(\partial^2+m^2)
-2g^{\lambda\sigma}g^{\rho\omega}(\partial^2+M^2) \\
&-&(\partial^\lambda\partial^\rho g^{\sigma\omega}+\partial^\sigma\partial^\rho g^{\lambda\omega}+
\partial^\lambda\partial^\omega g^{\sigma\rho}+\partial^\sigma\partial^\omega g^{\lambda\rho})\nonumber\\
&+&2(\partial^\lambda\partial^\sigma g^{\rho\omega}+\partial^\omega\partial^\rho g^{\sigma\lambda})
\nonumber
\end{eqnarray}

Several comments are in order here. A general symmetric tensor field
has ten 
components.
A massless spin two particle has only two degrees of freedom. The
tensor 
structure of the
kinetic term for the massless
tensor particles is therefore determined so that it should project out

two components out of
$G^{\mu\nu}$. In fact it is easy to check that the kinetic term in
Eq.~\ref{D}
in the massless case ($m^2=M^2=0$) is invariant under the four
parameter 
local gauge transformation
\begin{equation}
\delta G^{\mu\nu}=\partial_\mu\Lambda_\nu+\partial_\nu\Lambda_\mu
\label{gauge}
\end{equation}
These four gauge invariances together with four
corresponding gauge fixing conditions indeed eliminate
eight components of $G^{\mu\nu}$. This gauge invariance is broken by
the 
mass. However the
equations of motion even in the massive case lead to four constraints
\begin{equation}
\partial_\mu G^{\mu\nu}=0
\label{constr}
\end{equation}

This eliminates four degrees of freedom out of ten\footnote{In the 
interacting theory, 
Eq.~\ref{model}, the constraint equations are slightly modified, but 
they still provide four
conditions on the fields.}. In addition the scalar field $G^\mu_\mu$
decouples from the rest of the dynamics and can be neglected. In fact
we 
have added a parameter 
$M$ whose only purpose is to make $G^\mu_\mu$ arbitrarily heavy. All
in 
all therefore we are left
with five independent propagating degrees of freedom in $G^{\mu\nu}$,
which 
is the correct number
to describe a massive spin two particle.

The scalar glueball self interaction potential $V(\sigma)$ is fairly
general 
and the
following discussion will not depend on it. The natural choice to keep
in 
mind is a mass term
augmented with the standard
triple and quartic self interaction, although one could also consider
more 
complicated logarithmic
potential as is becoming a dilaton field.

Our first observation is that the Lagrangian Eq.~\ref{model} 
allows for static classical solutions
which are very reminiscent of the Abrikosov - Nielsen - Olesen vortices.
Consider a static field configuration of the form
\begin{eqnarray}
&&G^{ij}=G^{00}=0\nonumber \\
&&G^{0i}=G^{i0}\equiv a^i(\vec x)\nonumber \\
&&\sigma\equiv \rho(\vec x)-v
\end{eqnarray} 
For such a configuration the Lagrangian Eq.~\ref{model} reduces to
the Lagrangian of the Abelian Higgs model (AHM) in the unitary gauge.
This configuration is therefore 
a solution of equations of motion, provided $a^i$
and $\rho$ solve precisely the same equations as in the 
AHM with the only modification that the
scalar potential is given by $V(\sigma)$. We will call this 
configuration the tensor flux
tube (TFT). In fact it does carry a conserved tensor flux. To see this
let us consider the following operator
\begin{equation}
\tilde F^{\mu\nu\lambda}
=\epsilon^{\mu\nu\rho\sigma}\partial_\rho G_{\sigma\lambda}
\end{equation}
For single valued field $G_{\mu\nu}$ it satisfies the conservation equation
\begin{equation}
\partial_\mu \tilde F^{\mu\nu\lambda}=0
\label{cons}
\end{equation}
This is the analog of the homogeneous Maxwell equation of the
 AHM without monopoles
\begin{equation}
\partial_\mu \tilde F^{\mu\nu}=0
\end{equation}
For our static TFT solution we have
\begin{equation}
\tilde F_{i00}\equiv b_i=\epsilon_{ijk}\partial_j a_k
\end{equation}
and the flux through the plane perpendicular to the symmetry axis of TFT is
\begin{equation}
\Phi\equiv \int dS_i b_i=2\pi/g
\end{equation}
The energy per unit length of the TFT
is directly related to the masses of the two glueballs and to the 
inter glueball coupling constants $g^2$ and $v$. This quantity is a
natural candidate for
the string tension of the pure Yang - Mills theory.

There is one important proviso to this discussion.
As we noted the Lagrangian Eq.~\ref{model} reduces for TFT
configurations to the AHM Lagrangian in the unitary gauge. The
identification of the magnetic flux solutions in the unitary gauge is
a little tricky. Naively one would say that they can not exist, since
the vector potential is massive but for the magnetic fluxon it should
vanish at infinity as a power and not exponentially, 
since its contour integral does
not vanish even for contours taken infinitely far from the location of
the fluxon.
However one should remember that the mass term for the vector field in
unitary gauge
allows certain singular configurations.
Consider a vector potential configurations 
which far enough from the origin ($r_{perp}=0$) 
vanishes everywhere except on a half plain (let's say
$x=0, y>0$):
\begin{equation}
a_i=\Phi_z\delta_{1i}\delta(x)\theta(y)
\label{ai}
\end{equation}
It certainly describes a fluxon in the $z$ direction with the flux
$\Phi_z$.
The singularity in $a_i$
does not contribute to the energy, if the singularity is quantized. The
reason 
is that the vector field mass term
is strictly speaking not a simple $m^2a_i^2$.
It really is the remnant of the covariant derivative term of the Higgs
field
$|D_iH|^2$. This
expression when written in terms of the phase $\phi$ of the Higgs is 
$g^2\rho^2(A_i-{1\over g}\partial_i\phi_{mod 2\pi})^2$. It does not
feel jumps in the phase $\phi$ which are integer multiples of $2\pi$. 
Since the unitary gauge 
field $a_i$ is defined shifting $A_i$ by ${1\over g}\partial_i\phi$ 
the mass term in the AHM
Lagrangian is actually
\begin{equation}
m^2(a^i_{mod {2\pi\over g}\Delta})^2
\label{m}
\end{equation}
where $\Delta$ is the ultraviolet regulator - lattice spacing.
For smooth fields $a_i$ there is no difference between the two mass terms. 
However Eq.~\ref{m} admits singularities of the type of Eq.~\ref{ai}.
 provided 
\begin{equation}
\Phi_z={2\pi n\over g}
\end{equation}
with integer $n$.
For these values of the magnetic flux the lowest 
energy configuration has the form
$a_i=\Phi_z\delta_{1i}\delta(x)\theta(y)+b_i(x)$ with $b_i$ a smooth, 
exponentially decreasing
function. The energy per unit length of these configurations is
perfectly finite.

In our discussion of the TFT solutions in the effective glueball
theory
we have assumed that just like in 
the AHM the mass term (and the interaction terms) 
of the tensor field allows quantized discontinuities of the
form $2\pi\Delta$. This is not all that unnatural.
Our effective
theory can be thought of as a gauge theory with the ``spontaneously
broken'' gauge group of Eq.~\ref{gauge}. The analog of the phase field
$\phi$ 
in our model is played by the gauge parameter $\Lambda_\mu$ of 
Eq.~\ref{gauge}. All that is needed
for the discontinuities to be allowed is that $\Lambda_\mu$ (or at
least 
$\Lambda_0$) be a 
phase, or in other words for the gauge group to have a $U(1)$
subgroup. 
This does not seem to
be an extremely unnatural requirement, although to determine whether
this 
is indeed true
one would need to have some information about dynamics on distance
scales 
shorter than the inverse
glueball mass.

Continuing the same line of thought, we can identify the objects which
should play in this model
the same role as magnetic monopoles in the superconductor, that is the
objects that are confined by TFT.
The fields $G$ and $\sigma$ can couple
to much heavier objects, which from the low energy point of view are 
basically point like. At these
short distances Eq.~\ref{cons} should be modified, and we can consider 
the current
\begin{equation}
J^{\nu\lambda}=\partial_\mu \tilde F^{\mu\nu\lambda}
\label{mon}
\end{equation}
Again I stress that the current $J^{\nu\lambda}$ must have only very
high 
momentum components
$k>>m$, otherwise Eq.~\ref{model} will not describe faithfully the low 
energy sector of the theory.
Of course, this situation is very similar to AHM with heavy magnetic 
monopoles where the 
homogeneous Maxwell equation is also violated on the distance scales
of 
the order of the monopole
size. 
Now, since $\tilde F^{\mu\nu\lambda}$ is antisymmetric under the 
interchange of $\mu$ and $\nu$,
our newly born current is conserved
\begin{equation}
\partial_\nu J^{\nu\lambda}=0
\end{equation}
 The components of the current which are relevant to our TFT 
configuration are $J^{\mu 0}$.
Imagine that the underlying theory does indeed contain objects that 
carry the charge
$Q=\int d^3x J^{00}$.
Then the argument for confinement 
of these objects is identical to the argument for confinement of
magnetic 
monopoles
in AHM. 
If these objects are identified with heavy quarks this becomes
the 
picture of the QCD
confinement from the low energy point of view. 
Of course this last point is purely speculative and whether these
objects exist and have anything to do with quarks remains to be seen.

However what we learn from this discussion is that some kind of
confining objects appear under natural assumptions in the
effective theory. This is indeed reminiscent of the appearance of
confining QCD strings in 2+1 dimensions and gives us hope that the
whole approach makes sense.
What would be really interesting to see is how the spontaneously
broken $Z_N$ structure ties into this effective Lagrangian pattern. One would
hope that it should dictate the form of this Lagrangian as well as
the angular nature of the ``gauge'' parameter $\Lambda_\mu$ on a
similar level of rigor as in 2+1 dimensions. 

\section{Discussion.} 
Th main thesis of these notes is this. Confinement in gluodynamics can
be unambiguously characterized as spontaneous breaking of the magnetic
$Z_N$ symmetry. This symmetry breaking pattern under some mild
assumptions
also determines the structure of the low energy effective theory. 
The basic low energy degrees of freedom are magnetic vortices - the
objects that carry the magnetic $Z_N$ charge. Spontaneous breaking of
magnetic $Z_N$ is synonymous with condensation of magnetic vortices.
In
this effective theory confinement is manifested on the classical level
in a very simple and straightforward manner. In 2+1 dimensions this
picture is well substantiated. In 3+1 dimensions although some
elements are in place, there is still work to be done.
 
We note that a lot of numerical work has been done in the past several
years to relate the properties of magnetic vortices with confinement
\cite{greensite},\cite{tomboulis}. The properties of vortex
distributions seem to correlate very well 
with the confinement properties of the theory. This 
should be viewed as confirmation of the relation of the vortex
condensation to confinement.

There are many questions one can ask at this junction. Let me mention
her just two obvious ones. First, if at
zero temperature some global symmetry is spontaneously broken, one
expects it to be restored at some finite temperature $T_c$. Does this $Z_N$
symmetry restoring phase transition exist in gluodynamics? The
answer to this question is an emphatic yes. It has been shown in
\cite{kovner6} that the restoration of the magnetic symmetry is indeed
the deconfining phase transition of gluodynamics. The canonical order
parameter for the deconfinement phase 
transition therefore is the expectation value of the vortex
operator. In the high temperature deconfined phase 
in 2+1 dimensions $<V(x)>=0$, while in 3+1 dimensions
$<V(C)>\propto\exp\{-\alpha S\}$. This has been confirmed
numerically\cite{temperature}. 

Another interesting question is what happens to the magnetic symmetry
when the theory contains dynamical fundamental fields.
The answer to this is that the symmetry stays there, but now it does
not have a local order parameter. Physically this is because 
fundamentally charged particles have nontrivial Aharonov-Bohm phase
while scattering off a pointlike magnetic vortex created by $V$. So the 
properties of the $Z_N$ charge here are somewhat similar to the 
properties of the electric charge in QED - both are global symmetries
and both do not have a local order parameter\footnote{Recall that any
physical gauge invariant operator that carries electric charge is
nonlocal, since it creates a long range electric field.}. Indeed it can
be shown\cite{kovner7} that the $Z_N$ symmetry in low energy effective theory
has to be gauged. This leads to some interesting dynamics and allows
one to construct a bag like picture of baryons\cite{kovner7}.

Still in my mind the most interesting question is how to extend
intelligently this picture to 3+1 dimensions and how to relate the
magnetic symmetry structure with the expected form of the effective
low energy Lagrangian. This ought to be the direction of the main
effort in this area.

\section*{Acknowledgments}
This work was supported by PPARC.
I am especially grateful to B. Rosenstein for long term enjoyable
collaboration which lead to my understanding of most of the ideas presented
here. I also enjoyed collaborating with C. Korthals Altes and C. Fosco
and am indebted to I. Kogan, B. Lucini  and M. Teper for many interesting and
stimulating discussions.

\end{document}